\pdfoutput=1

\documentclass[fleqn,usenatbib]{mnras}

\usepackage[T1]{fontenc}
\usepackage{ae,aecompl}

\usepackage{graphicx}	
\usepackage{amsmath}	
\usepackage{amssymb}	
\usepackage[dvipsnames]{xcolor}
\usepackage{lscape}
\usepackage{rotating}
\usepackage{multirow}

\usepackage{newtxtext,newtxmath}

\newcommand{\uu}{`\=U`\=u}

\title[NESS Overview paper]{The Nearby Evolved Stars Survey II: Constructing a volume-limited sample and first results from the James Clerk Maxwell Telescope}

\author[P. Scicluna et al.]{P. Scicluna,$^{1, 2}$\thanks{E-mail: peterscicluna@asiaa.sinica.edu.tw} 
F. Kemper,$^{3,1}$
I. McDonald,$^{4,5}$
S. Srinivasan,$^{6,1}$
A. Trejo,$^{1}$
S.~H.~J. Wallstr\"om,$^{7,1}$\newauthor
J.~G.~A. Wouterloot,$^{8}$
J. Cami,$^{9,10,11}$
J. Greaves,$^{12}$
Jinhua He,$^{13,14,15}$ 
D. T. Hoai,$^{16,17}$
Hyosun Kim,$^{18}$ \newauthor
O.~C. Jones,$^{19}$
H. Shinnaga,$^{20,21}$
C.~J.~R. Clark,$^{22}$
T. Dharmawardena,$^{1,23}$
W. Holland,$^{19}$ 
H. Imai,$^{21,24}$\newauthor 
J. Th. van Loon,$^{25}$
K.~M. Menten,$^{26}$
R. Wesson,$^{27}$
H. Chawner,$^{12}$
S. Feng,$^{28}$
S. Goldman,$^{22}$ 
F.C. Liu,$^{29}$ \newauthor
H. MacIsaac,$^{9,10}$
J. Tang,$^{28, 30}$
S. Zeegers,$^{1}$ 
K. Amada,$^{20}$ 
V. Antoniou,$^{31, 32}$ 
A. Bemis,$^{33}$
M.~L. Boyer,$^{22}$ \newauthor
S. Chapman,$^{34}$
X. Chen,$^{35}$
S.-H. Cho,$^{18}$
L. Cui,$^{36}$ 
F. Dell'Agli,$^{36}$
P. Friberg,$^{8}$
S. Fukaya,$^{20}$ 
H. Gomez,$^{12}$ \newauthor
Y. Gong,$^{26}$
M. Hadjara, $^{14,15}$
C. Haswell,$^{5}$
N. Hirano,$^{1}$
S. Hony,$^{38}$ 
H. Izumiura,$^{37}$
M. Jeste,$^{26}$
X. Jiang,$^{40}$\newauthor
T. Kaminski,$^{32}$ 
N. Keaveney,$^{41}$
J. Kim,$^{18,35}$ 
K.~E. Kraemer,$^{42}$ 
Y.-J. Kuan,$^{29,1}$ 
E. Lagadec,$^{43}$
C.F. Lee,$^{1}$\newauthor
D. Li,$^{44,45,46}$
S.-Y. Liu,$^{1}$ 
T. Liu,$^{35}$ %
I. de Looze,$^{47,27}$
F. Lykou,$^{48,49}$ 
C. Maraston,$^{50}$  
J.~P. Marshall,$^{1,51}$\newauthor
M. Matsuura,$^{12}$
C. Min,$^{52}$ 
M. Otsuka,$^{53,1}$ 
M. Oyadomari,$^{20}$
H. Parsons,$^{8}$
N.~A. Patel,$^{32}$
E. Peeters,$^{9,10,11}$ \newauthor
T.~A. Pham,$^{16}$
J. Qiu,$^{54,33}$ 
S. Randall,$^{3}$ 
G. Rau,$^{55,56}$ 
M.~P. Redman,$^{41}$ 
A.~M.~S. Richards,$^{4}$
S. Serjeant,$^{5}$\newauthor
C. Shi,$^{57,58}$
G.~C. Sloan,$^{22,59}$ 
M.~W.~L. Smith,$^{12}$
J.~A. Toal\'{a},$^{6}$ 
S. Uttenthaler,$^{60}$  
P. Ventura,$^{36}$
B. Wang,$^{44}$\newauthor
I. Yamamura,$^{61,62}$
T. Yang,$^{36}$ 
Y. Yun,$^{18}$
F. Zhang,$^{13}$
Y. Zhang,$^{63,49}$ 
G. Zhao,$^{44}$
M. Zhu$^{44}$
and A.~A. Zijlstra$^{4}$
\\
Affiliations are listed at the end of this paper
}
\date{Accepted XXX. Received YYY; in original form ZZZ}

\pubyear{2019}

\begin{document}
\label{firstpage}
\pagerange{\pageref{firstpage}--\pageref{lastpage}}
\maketitle
\begin{abstract}
The Nearby Evolved Stars Survey (NESS) is a volume-complete sample of $\sim$850 Galactic evolved stars within 3\,kpc at (sub-)mm wavelengths, observed in the CO $J = $ (2--1) and (3--2)  rotational lines, and the sub-mm continuum, using the James Clark Maxwell Telescope and Atacama Pathfinder Experiment. NESS consists of five tiers, based on distances and dust-production rate (DPR). We define a new metric for estimating the distances to evolved stars and compare its results to \emph{Gaia} EDR3. Replicating other studies, the most-evolved, highly enshrouded objects in the Galactic Plane dominate the dust returned by our sources, and we initially estimate a total DPR of $4.7\times 10^{-5}$ M$_\odot$ yr$^{-1}$ from our sample. Our sub-mm fluxes are systematically higher and spectral indices are typically shallower than dust models typically predict. The 450/850 $\mu$m spectral indices are consistent with the blackbody Rayleigh--Jeans regime, suggesting a large fraction of evolved stars have unexpectedly large envelopes of cold dust.
\end{abstract}

\begin{keywords}
surveys -- catalogues -- stars: AGB and post-AGB -- stars: mass-loss -- stars: winds, outflows
\end{keywords}



\section{Introduction}

The asymptotic giant branch (AGB) represents the terminal  evolutionary stage of low- to intermediate-mass stars ($M \lesssim 8$ M$_\odot$), before they become white dwarfs potentially surrounded by \mbox{(pre-)}planetary nebulae. Particularly in the near- and mid-infrared, AGB stars dominate light from galaxies with intermediate-age and old populations \citep[e.g.][]{Maraston+06,Melbourne2012,Riffel2015}, while their ejecta dominates the evolution of light elements (primarily C and N) and many of the $s$-process elements in our present-day Galaxy \citep[e.g.][]{KarakasLattanzio14}.

However, a complex interplay of different physical mechanisms makes the AGB a challenging phase of evolution to model. Thermal pulses occur as helium burning repeatedly ignites in conditions of thermal instability, allowing convection to dredge carbon-rich matter up from near the degenerate core to the stellar surface \citep[Third Dredge Up (3DU), e.g.][]{Herwig05}. Repeated dredge-up episodes gradually increase the carbon content in the stars' atmospheres and envelopes, leading to the formation of carbon stars. However, if hot-bottom burning occurs at the dredge-up site, carbon is transmuted into nitrogen, and the star remains oxygen-rich \citep[e.g.][]{KarakasLattanzio14}.

Near-simultaneously, AGB stars become unstable to surface pulsations. These can levitate material from the stellar photosphere to altitudes where molecules can condense into dust.
Most of the carbon and oxygen binds into CO, leaving an oxygen-dominated chemistry around most stars, but a carbon-dominated chemistry around carbon stars. The levitated molecules then go on to form oxygen-rich or carbon-rich dust. Radiation pressure on this dust forces it from the star, and collisional coupling with the surrounding gas drives an enhanced stellar wind, with a mass-loss rate much greater ($\dot{M} \sim 10^{-7}$ to $10^{-4}$ M$_\odot$ yr$^{-1}$; e.g. \citealt{Danilovich15}) than at earlier evolutionary phases \citep[e.g.][]{HofnerOlofsson18}.

The primary unknowns are the 3DU efficiency and the mechanisms determining $\dot{M}$.
Since $\dot{M}$ for thermally pulsating AGB stars generally exceeds the rate of hydrogen burning ($\dot{M}_{\rm H} \sim 10^{-7}$ to $10^{-6}$ M$_\odot$ yr$^{-1}$; e.g., \citealt{Marigo08}), $\dot{M}$ dictates the star's longevity and the amount of material that is dredged up. Simultaneously, the wind chemistry dictates the dust mass and opacity, hence whether radiation pressure can drive a wind. The two problems are therefore strongly interlinked \citep[e.g.][]{LagadecZijlstra08, Uttenthaler2019}.

The material ejected by evolved stars contributes to subsequent star formation, shaping the chemical evolution of galaxies.
Thousands of dusty evolved stars have been identified in Local Group galaxies with {\it Spitzer}, highlighting the roles of various sub-populations \citep[e.g.][]{Boyer11,Boyer15,Boyer17,DellAgli16,Britavskiy15,Meixner06}.
These galaxies can be studied globally, but gas tracers (including masers; \citealt{Goldman17}) are only detectable from the very brightest stars \citep{Groenewegen16,Matsuura16}, for the remainder we can only study the dust.
Conversely, molecular lines from evolved stars in the solar neighbourhood can be detected even with modest telescopes. 
However, Galactic surveys must contend with interstellar extinction, significant confusion, the entire range of on-sky directions, variations of angular size, and high dynamic range.

Collectively, these uncertainties can dramatically affect galaxy evolution.
Failing to quantify and characterise mass loss properly prevents the development of predictive models for population synthesis and galaxy evolution.
For example, \citet{Pastorelli19} covered a range of mass-loss and 3DU prescriptions for the Small Magellanic Cloud (SMC), showing their large impact on the inferred population.
Similar calibrations for metal-rich environments, relevant to galaxies such as our own, do not exist, although spectral signatures of AGB stars have been detected in such environments both locally \citep[e.g.][]{Boyer2019} and out to high redshift \citep[e.g.][]{Maraston+06}.

This paper introduces the Nearby Evolved Stars Survey\footnote{\url{http://evolvedstars.space}} (NESS): a volume-complete, statistically representative set of AGB stars in the solar neighbourhood.
We describe the formulation, objectives, sample selection (Sect.~\ref{sec:form}) and observing strategy in Sect.~\ref{sec:obs}. The sample itself is explored in Sect.~\ref{sec:obj}, with some first results of the survey presented in Sect.~\ref{sec:sum}.
The raw data will be publicly available on the Canadian Astronomy Data Centre (CADC), while any scripts required to produce plots and tables are available in a GitHub repository\footnote{\url{https://github.com/evolvedstars/NESS_02_OverviewPaper}}. 
The full versions of tables presented in this paper are available on the NESS website and through Vizier.

\section{Survey design}\label{sec:form}

\subsection{Motivation for a sub-millimetre survey of evolved stars}\label{sec:comotivate}

\subsubsection{Available methods}\label{sec:comethods}

Evolved stars are typically analysed using two tracers: dust and molecular lines. 
NESS aims to build a coherent picture of mass-loss rates, by comparing these two mass tracers.

Dust-based mass-loss rates scale with the fraction of stellar light reprocessed from the optical to the mid-infrared (and derive the optical depth). This traces warm dust ($\sim$50--1500 K) close to the star. The scaling factor depends on the distance and luminosity of the star, and requires assumptions on the expansion velocity (e.g., from CO lines or OH masers) of the wind and its radial profile, how strong the dynamical coupling between dust and gas is, the wind's geometry and clumpiness, the dust-to-gas ratio, and various geometrical, chemical and mineralogical properties of the dust grains \citep[e.g.][]{vanLoon2006,McDonald11,Jones14}, as these factors all influence how stellar radiation is reprocessed by the outflowing dust.

By contrast, gas mass-loss rates from radio or (sub)millimetre wavelength observations can directly yield the wind's velocity and density structure. This simplifies the scaling factors, requiring only the unknown but modellable excitation conditions of the observed molecule and its abundance in the wind. 
Of the abundant molecules, CO is expected to be the most stable, with a consistent abundance of $\sim$10$^{-4}$ with respect to molecular hydrogen across a range of mass-loss rates and chemistry \citep[e.g.][]{Schoier02,Ramstedt08,DeBeck10,Danilovich15}, though it will be lower in metal-poor environments \citep[e.g.][]{Leroy11}.

The gaseous envelope ejected over the last few millennia is cold (kinetic temperatures are a few $\times$ 10 K). With energies of $E_u/k = 5 - 50$\,K, \citep{Pickett1998JPL}, low-$J$ pure-rotational transitions of CO are sensitive to the molecular content of the entire cold envelope, particularly the outer regions \citep{Kemper03}. Many of these transitions fall in the (sub-)mm atmospheric windows, facilitating ground-based observation.
Timescales of $\sim$10\,000 years can be probed \citep{Mamon88, Groenewegen17, Saberi2019}, and time variability in mass-loss rates can sometimes be uncovered \citep[e.g.][]{Olofsson90, Decin06,Guelin2018,Maercker10,Dharmawardena2019}. 

To generate dust-based estimates comparable to $\dot{M}$ from CO lines, we must therefore probe the bulk of the dust mass, which is likely to be at similarly cold temperatures to the gas seen in low-excitation CO lines.
The sub-mm continuum traces emission from the coldest dust, at similar temperatures as the low-$J$ CO lines.

\subsubsection{Context of existing studies}\label{sec:costudies}

CO-based mass-loss rates have successfully been exploited by a wide range of studies. However, these are all either limited in scope or exhibit biases that make comparative statistics difficult.

Many existing studies focus on the brightest AGB stars with the highest mass-loss rates. ``Extreme'' mass-losing stars often dominate the dust budget of galaxies \citep[e.g.][]{Riebel12,Boyer12,Srinivasan16}, and are thus observed to explore these rapid and catastrophic evolutionary phases, and to constrain the dust budget of galaxies. However, strong mass loss ($\dot{M} \gtrsim 10^{-6}$ M$_\odot$ yr$^{-1}$) only occurs during the final stages of the AGB evolution of higher-mass AGB stars, hence observing them misses the mass loss that occurs in lower-mass AGB stars, or earlier on in AGB evolution, frustrating efforts to quantify these final stages in evolutionary models and biasing our understanding of chemical enrichment. In particular, very few AGB stars have been observed in the early phases of their dusty wind production, or before strong dust-production starts \citep{Groenewegen14,Kervella16,McDonald16,McDonald18}, despite these being both numerically the most numerous AGB stars in most galaxies \citep[e.g.][]{Boyer15,Boyer15b} and the terminal phase of the AGB for low-mass stars \citep[e.g.][]{McDonald15a}.

Other samples offer volume completeness, but only of specific types of targets, such as only carbon stars \citep{Olofsson93} or S-type stars \citep[where C/O $\sim$ 1; e.g.,][]{Sahai95,GroenewegenDeJong98,Ramstedt09}.
These samples can be compared \citep[e.g.][]{Ramstedt09}, but the different relative sizes, completenesses and volumes of these samples make it difficult to draw robust comparative statistics.

Most of these studies are relatively modest, comprising typically of tens of stars \citep[e.g.][]{Neri98,Kemper03,Teyssier06,McDonald18,Dharmawardena2018}, and studies with larger samples of up to $\sim$100 stars \citep[e.g.][]{KerschbaumOlofsson99,Olofsson02,Loup93,Kastner93, Zuckerman1986,Zuckerman1986b,Zuckerman1986c,Zuckerman1989} typically have tens of detections. \citet{Nyman92} detected some 160 out of $\sim$500 IRAS-identified evolved stars in CO $J$=1--0, but did not target a complete sample or full sky coverage, and had relatively low detection rates due to a lack of sensitivity; while \citet{Nyman92} had $T_{\rm sys} \sim 400 - 1000\,$K, the JCMT routinely observes with $T_{\rm sys} \lesssim100\,$K enabling a factor of 3 or more improvement in sensitivity. 
This  made it difficult for previous studies to accurately extract trends across the complex parameter space of AGB evolution, and to be robust against statistical outliers. This motivates the need for a volume-complete survey of nearby AGB stars, of sufficiently large scale to extract relationships that can improve comparative and evolutionary studies.

The optimal observational setup depends strongly on the spatial scale to be recovered. While the CO $J$=1--0 line has historically been the easiest to observe \cite[e.g.][]{Nyman92}, the large beam sizes and resultant contamination by cold interstellar gas and dust mean that the $^{12}$CO $J$=2--1 and 3--2 transitions (230.538 and 345.796 GHz, respectively) and sometimes the related $^{13}$CO transitions (220.399 and 330.588 GHz, respectively) are generally chosen to observe AGB stars (e.g. Fig.~\ref{fig:1213}). Higher spatial resolution can be obtained with interferometers, but pressure on these instruments prohibits very large programmes at high sensitivity, and emission can be filtered out at large scales due to incomplete filling of the $uv$ plane. Large single-dish telescopes provide high sensitivity without filtering out emission.
In particular, the James Clerk Maxwell Telescope (JCMT, 15 metre diameter, with resolutions of 21$^{\prime\prime}$/ 14$^{\prime\prime}$ at 230/345 GHz, respectively) and Atacama Pathfinder EXperiment (APEX, 12 metre diameter, 25$^{\prime\prime}$/ 17$^{\prime\prime}$), equipped with multi-beam detector arrays, provide high mapping speeds capable of observing hundreds of stars, efficiently revealing historic mass loss through extended emission, while the observing methods employed ensure that all the flux is captured. 

By making this selection, the NESS project is performing a wide survey of a large number of AGB stars, $\sim$850, albeit at the cost of recovering limited spatial information about the envelope of each one. This places NESS as the widest project, with strong synergies to other, ongoing large observing programmes. At medium spatial resolutions (5$\farcs$5/3--4$^{\prime\prime}$) with only slight expense of $uv$ coverage (maximum recoverable scale, MRS $\sim$ 25$^{\prime\prime}$/18$^{\prime\prime}$) and sample size ($\sim$180 stars), is the Morita (Atacama Compact Array) survey DEtermining Accurate mass-loss rates for THermally pulsing AGB STARs (DEATHSTAR; \citealt{Ramstedt20}\footnote{ \url{http://www.astro.uu.se/deathstar}.}). Finally, at the highest resolutions (0.05--0.025$^{\prime\prime}$/0.03--0.016$^{\prime\prime}$) of the main Atacama Large Millimeter/submillimeter Array (ALMA), is the project ``ALMA Tracing the Origins of Molecules forming dust In Oxygen-rich M-type stars'' (ATOMIUM; \citealt{Decin20}\footnote{\url{https://fys.kuleuven.be/ster/research-projects/aerosol/atomium}.}), which includes 14 nearby AGB stars and red supergiants.

\subsection{Survey aims}\label{sec:aims}

NESS aims to: (i) study the rate and properties of the enriched matter returned to the ISM by evolved stars; and (ii) explore the physics of dust-laden stellar winds, particularly their onset and time evolution,
through the following objectives.

\begin{description}
\item[{\it The statistics of local evolved stars.}]
Our homogeneous, volume-complete sample of AGB stars allows largely unbiased relationships between observables. To benefit, we aim to generate homogeneous parameters for our sample.

\item[{\it The return of enriched material to the Solar Neighbourhood.}]
The initial-to-final mass relation for stars is relatively well-known, but the fraction lost via a dust-laden, chemically enriched AGB wind is not. Hence, the dust-injection rate returned to the Milky Way ISM is poorly determined \citep[e.g.][]{JuraKleinmann89}. In parallel with wider analysis of all-sky surveys (Trejo et al., \emph{in prep.}) and using chemical, morphological and other calibrative input from other surveys like DEATHSTAR and ATOMIUM, the volume-complete, tiered nature of NESS aims to calibrate the contribution to the ISM by evolved stars at different stages along the AGB evolution, including better calibration of the relative contributions of carbon- and oxygen-rich stars  (by including sources further from the Sun than previous studies, hence obtaining a fairer picture of our Galactic environment), and better statistics on lower-luminosity and lower mass-loss rate objects that have been missed by previous studies.

\item[{\it The dust-to-gas ratio in AGB winds.}]
The best systematic study of this ratio to date \citep{Knapp1985b} had low sensitivity to both high and low $\dot{M}$ sources. NESS specifically targets such sources and infers whether variations in the dust-to-gas ratio exist among stars of different evolutionary phase and chemical type.

\item[{\it $^{13}$CO / $^{12}$CO abundances.}]
\begin{figure}
    \centering
    \includegraphics[width=0.485\textwidth, clip=true, trim=0.5cm 0.25cm 1.5cm 1.45cm]{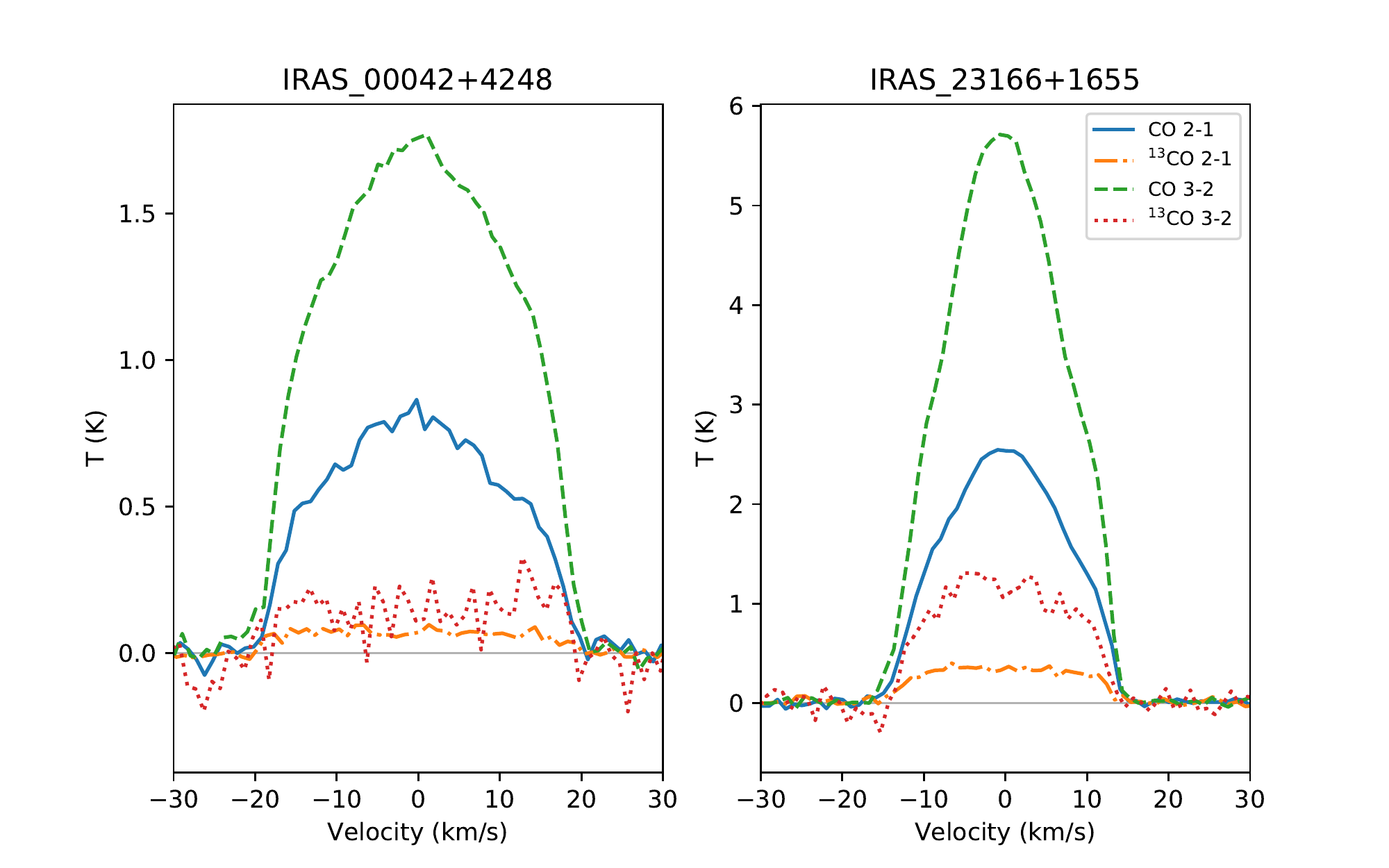}
    \caption{Examples comparing main-beam temperatures of $^{12}$CO and $^{13}$CO lines for two sources: KU~And (left) and AFGL3068 (right), both of which are in Tier 3. All four lines are clearly detected at high significance.}
    \label{fig:1213}
\end{figure}

The $^{13}$C/$^{12}$C ratio probes dredge-up efficiency \citep[e.g.][]{GreavesHolland1997}. 
NESS sample will provide an order-of-magnitude increase in the number of AGB stars for which $^{13}$CO observations have been made \citep[cf.][]{GreavesHolland1997,DeBeck10,Ramstedt14}.  The newly observed sources will largely comprise of M-type stars, as these dominate AGB stars in the solar neighbourhood. The 3DU process in such stars has not yet been sufficiently active to enhance their C/O ratios close to unity, when they become observationally S-type stars showing ZrO bands \citep[e.g.][]{Smolders12}. This not only allows us to explore the poorly quantified effect on $^{13}$C of earlier mixing processes \citep[notably mixing on the upper RGB and second dredge-up, e.g.,][]{KarakasLattanzio14}, but potentially allows us to do so in a mass-dependent way, by linking stars to evolutionary models of specific masses \citep[e.g.][]{Trabucchi21}. Through this process, and in combination with other tracers such as Tc \citep[e.g.][]{Uttenthaler2019}, we can also hope to probe the onset of 3DU among AGB stars.

\item[{\it A new mass-loss law for solar-metallicity AGB stars.}]

The physical complexity of mass loss means empirical formulae are used in stellar-evolution models \citep[e.g.][]{Dotter08,Paxton11,Bressan12}. These have limited ranges of validity and suffer from systematic uncertainties, caused by small sample sizes, the accuracy of the input stellar properties, or the conversion from dust opacity to $\dot{M}$. NESS covers the entire gamut of dust-producing AGB stars found in the solar neighbourhood, and directly measures $\dot{M}$ from gas tracers. We can more directly relate $\dot{M}$ to observable stellar parameters, and link it with those at sub-solar \citep[e.g.][]{Boyer15,McDonaldZijlstra15,McDonaldTrabucchi19} and super-solar \citep[e.g.][]{vanLoon08,Miglio12} metallicities, to provide a single law for stellar evolution modellers.

\item[{\it Spatially-resolved mass loss and irradiation of AGB envelopes.}] Temporal and geometric variations in mass loss can be traced by mapping the distribution of ejecta around AGB stars. Irradiation of envelopes dissociates CO \citep{Mamon88}, so mapping the extent of the CO and dust envelopes allows NESS to reveal how ejecta of stars is influenced by their surroundings.

\item[{\it Revealing cold dust.}] Cold dust represents a key probe of the historic mass loss and the properties of the dust; see Sect.~\ref{sec:cold}. Mapping sources in the sub-mm continuum allows NESS to reveal this dust for a wide variety of sources and explore the evolution of AGB dust grains as they enter interstellar space.

\end{description}

\subsection{Sample selection}\label{sec:samp}

\subsubsection{Determining distances}\label{sec:distances}

We require a large, volume-complete sample of Galactic evolved stars covering the largest possible range of $\dot{M}$, homogeneously observed in both CO lines and continuum, aiming to minimise biases. 
Defining the sample volume depends critically upon our ability to determine distances. Unfortunately, while distance estimates are rapidly improving, distances to Galactic evolved stars remain poor. We must make assumptions, and revisit the volume-completeness of the NESS survey in future, ultimately allowing us to complete our survey aims.

Where observed, maser parallax provides the most exquisite precision, but only evolved stars with the highest $\dot{M}$ exhibit the masing transitions required. We have included maser distances as our preferred choice where available \citep[e.g.][]{Orosz2017}. 

Convective motions and dust obscuration can move the astrometric centres of AGB stars in the optical, introducing significant astrometric noise to optical parallax measurements \citep[e.g.][]{Chiavassa18}. Similarly, AGB stars are poorly captured by the prior model used in \citet{Bailer-Jones18}. Consequently, data from \emph{Gaia} DR2 \citep{GaiaDR2} were not used (\emph{Gaia} EDR3 is addressed below), as they rely solely on short-timescale \emph{Gaia} data (see \citealp{McDonald18} for further discussion). Instead, we adopt the parallaxes from the \emph{Tycho--Gaia} Astrometric Solution (TGAS) of \emph{Gaia} Data Release 1 (DR1) \citep{GaiaDR1} as our next preference, followed by the \emph{Hipparcos} parallaxes \citep{vanLeeuwen07}, provided their fractional uncertainty is small ($\sigma_\varpi / \varpi < 0.25$).

Other distance-determination methods typically only apply to a small subset of AGB stars; e.g., period--luminosity relationships can only be used if the pulsation mode and mean magnitude are known \citep[e.g.][]{Uttenthaler13, Huang2018, Yuan2018, Goldman2019}, as overlap between different sequences and scatter from single-epoch photometry can confuse measurements.

Instead, we take a statistical approach to missing distances, using the luminosity distribution of evolved-star candidates in the Large Magellanic Cloud (LMC), taken from the \emph{Spitzer} Surveying the Agents of Galactic Evolution (SAGE) database \citep{Meixner06}, accounting for the geometry and thickness of the LMC.
The luminosity of each star is determined by integrating its SED from the optical to the far-infrared and the luminosities of all sources are combined to give a probability distribution, from which we extract the median and the central 68 per cent.
This gives a ``typical'' luminosity and uncertainty of $6200^{+2800}_{-3900}$\,L$_\odot$ for the LMC population. 

Assuming that AGB stars in the solar neighbourhood have sufficiently similar luminosities to stars in the LMC, we can apply this same luminosity distribution to derive a probabilistic distribution of distances to individual Galactic sources.
A galaxy's luminosity distribution is mainly set by its star-formation history, as the final AGB luminosity depends on initial mass. Secondary effects include variation in mass loss, stellar evolution and the initial-mass function with metallicity.
To test this assumption, we compare the AGB luminosity functions of the LMC and SMC to each other, and to the solar neighbourhood. No significant difference is seen (e.g., Fig.\ 15 in \citealt{Srinivasan16}). The spread in median luminosity, 17 per cent \citep{Boyer15,McDonald17}, indicates a systematic galaxy-to-galaxy error of 8.5 per cent in this distance-based method.
In comparison, the intrinsic width of the luminosity distribution corresponds to a distance uncertainty of $\approx$25 per cent.
Consequently, we expect any galaxy-to-galaxy differences to produce much smaller systematic uncertainties than the random uncertainties inherent in the measurement itself (see also Section \ref{sec:gaiaEDR3}).
We therefore use this LMC luminosity distribution and photometry from infrared all-sky surveys to determine distances to Galactic sources where parallaxes from optical or maser astrometry are unavailable or imprecise.

Such a luminosity-based, volume-limited survey is subject to forms of Malmquist bias \citep{Malmquist25}, whereby many sources from larger distances scatter into our distribution, while only a few sources from smaller distances scatter out. The $\dot{M}$ cutoffs that define our tiers (below) further re-enforce this bias, ensuring distant objects scattering into the sample are assigned anomalously high mass-loss rates, hence are placed into more extreme categories of mass loss. Furthermore, parallax errors generate a Lutz--Kelker bias \citep{LutzKelker73}, preferentially scattering sources into our sample for the same reasons. Hence, while the distances to our sources undergo frequent revision, we expect our samples to be relatively robust against obtaining larger datasets in future, though ultimately some sources may no longer meet our distance criteria. A small Malmquist bias in the LMC sample cancels out some of these effects, but AGB stars in the SAGE data are bright enough that this is largely negligible. In the following sections, we attempt to reduce these biases in our data.

\subsubsection{Object selection}\label{sec:objects}

The \emph{Infrared Astronomical Satellite} (\emph{IRAS}) point-source catalogue \citep{Beichman88} is our fiducial reference for source selection. \emph{IRAS} photometry were matched against the 2MASS catalogue \citep{Cutri20032mass} to find the nearest neighbour within 30$\arcsec$, providing homogeneous near- and mid-infrared photometry for all sources. 
The photometry were then integrated numerically to estimate the bolometric fluxes of the sources; we then compute the distribution of distances as that required to scale the bolometric flux to the LMC luminosity distribution. The 50th centile luminosity (6200 L$_\odot$) is chosen as representative, so that we have a point-distance estimate for each source following the constraints described below in Sects.~\ref{subsec:staring}\,\&\,\ref{subsec:mapping}. In order to restrict their sample to stars with mass-loss rates $>2\times 10^{-6}$ M$_\odot$ yr$^{-1}$, \citealt{JuraKleinmann89} select sources with a minimum \emph{IRAS} 60~$\mu$m flux of 10 Jy at 1 kpc. In their work as in ours, the mass-loss rates are computed from the dust-production rates assuming a gas-to-dust ratio of about 200. Their limit therefore corresponds to a DPR of $1\times 10^{-8}$ M$_\odot$ yr$^{-1}$. Extending this criterion to 2 kpc, we select sources brighter than 2.5 Jy, thus reducing the full point-source catalogue to $\sim$60,000 sources. This removes most extra-galactic sources and less-evolved stars, and is set sufficiently high that it avoids strong effects from Malmquist bias. We consider the effects of this choice on completeness in Section \ref{sec:completeness}.

To this sample, we add evolved stars within 300 pc from \citet{McDonald12} and \citet{McDonald17}, which would otherwise have been excluded by the IRAS\,60\,$\mu$m flux cut mentioned above. Stars were added if they have $T_{\rm eff} < 5500$ K and $L > 700$ L$_\odot$, and if their \emph{Hipparcos} or TGAS parallaxes have fractional uncertainties of $\sigma_\varpi / \varpi < 0.25$ (560 sources meet these criteria). This cutoff matches the expected accuracy of our luminosity-based distances and avoids strong effects from Lutz--Kelker bias.

Since $\sim$60,000 sources are beyond our capability to observe, we design a tiered system to observe representative samples of them, selected from a plane of DPR versus distance, using first-order estimates of the DPR outlined in Sect.~\ref{sec:tiers}, and cuts outlined in Sect.~\ref{subsec:staring}. This results in a set of 2277 potential targets, from which sources with {\sc simbad}\footnote{\url{http://simbad.u-strasbg.fr/}} classifications that are inconsistent with dusty evolved stars are removed\footnote{To measure mass return, we are primarily interested in sources with non-negligible ongoing or past mass loss, i.e., for low- and intermediate-mass stars, those on the early-AGB up to and including planetary nebulae (PNe), and for high-mass stars red supergiants (RSGs), yellow hypergiants (YHGs), B[e] supergiants (B[e]SGs), luminous blue variables (LBVs) and Wolf-Rayet stars (WRs).}. Of the 852 remaining sources in our final sample (defined below), only nine have maser distances available, and seven and 193 respectively have TGAS and \emph{Hipparcos} measurements with uncertainties $< 25$ per cent. All remaining sources use luminosity distances.

\subsubsection{Determining preliminary mass-loss rates}\label{sec:tiers}

We fit the matched photometry in the 2MASS $J$, $H$, and $K_s$ bands and the \emph{IRAS} 12, 25 and 60 $\mu$m bands with models from the Grid of Red supergiant and AGB ModelS (GRAMS; \citealt{Sargent2011,Srinivasan2011}) and extract dust-production rates ($\dot{M}_{\rm dust}$) for the entire sample. 
$\dot{M}_{\rm dust}$ is thus used as a criterion to define a tiered survey, limited in distance and $\dot{M}_{\rm dust}$ as described in Sect.\ \ref{subsec:staring}. Our source lists are additionally divided into two groups, a large set to be observed in spatially unresolved modes (the ``staring'' sample) and a smaller group to be mapped in detail (the ``mapping'' sample).

GRAMS is tailored to the LMC, and the assumed dust properties reflect its metal-paucity \citep[i.e. the relative abundance of different dust species may be marginally different in Galactic sources; e.g.,][]{Srinivasanetal2010}.
However, for the broadband photometry used here, this should provide sufficient precision to perform sample selection. 
Under certain assumptions of velocity scaling, dust opacity and dust-to-gas ratio, $\dot{M}_{\rm dust}$ should provide a close proxy to total mass-loss rate, $\dot{M}$: an assumption that NESS will test (Section \ref{sec:obj}).
While GRAMS also returns a best-fit chemical classification (O- or C-rich), the seven bands of photometry used in this paper are insufficient to accurately constrain the dust chemistry. Given that M-stars are vastly more numerous in the Milky Way, we therefore assume that each source is oxygen-rich \emph{unless} there is spectroscopic confirmation of C-rich dust chemistry from either the \emph{IRAS} LRS or \emph{ISO} SWS spectra, with the \emph{ISO} SWS classification from \citet{Kraemer2002ApJS..140..389K} taking precedence over \emph{IRAS} LRS classifications from \citet{Kwok1997ApJS..112..557K}. This assumption applies only to the determination of initial DPRs and is necessary because the choice of dust chemistry can make a significant difference to the derived DPR. We demarcate these unclassified sources separately below.

\subsubsection{Staring sample}
\label{subsec:staring}

\begin{figure}
    \centering
    \includegraphics[width=0.475\textwidth, clip=true, trim=0.5cm 0.25cm 1.5cm 1.45cm]{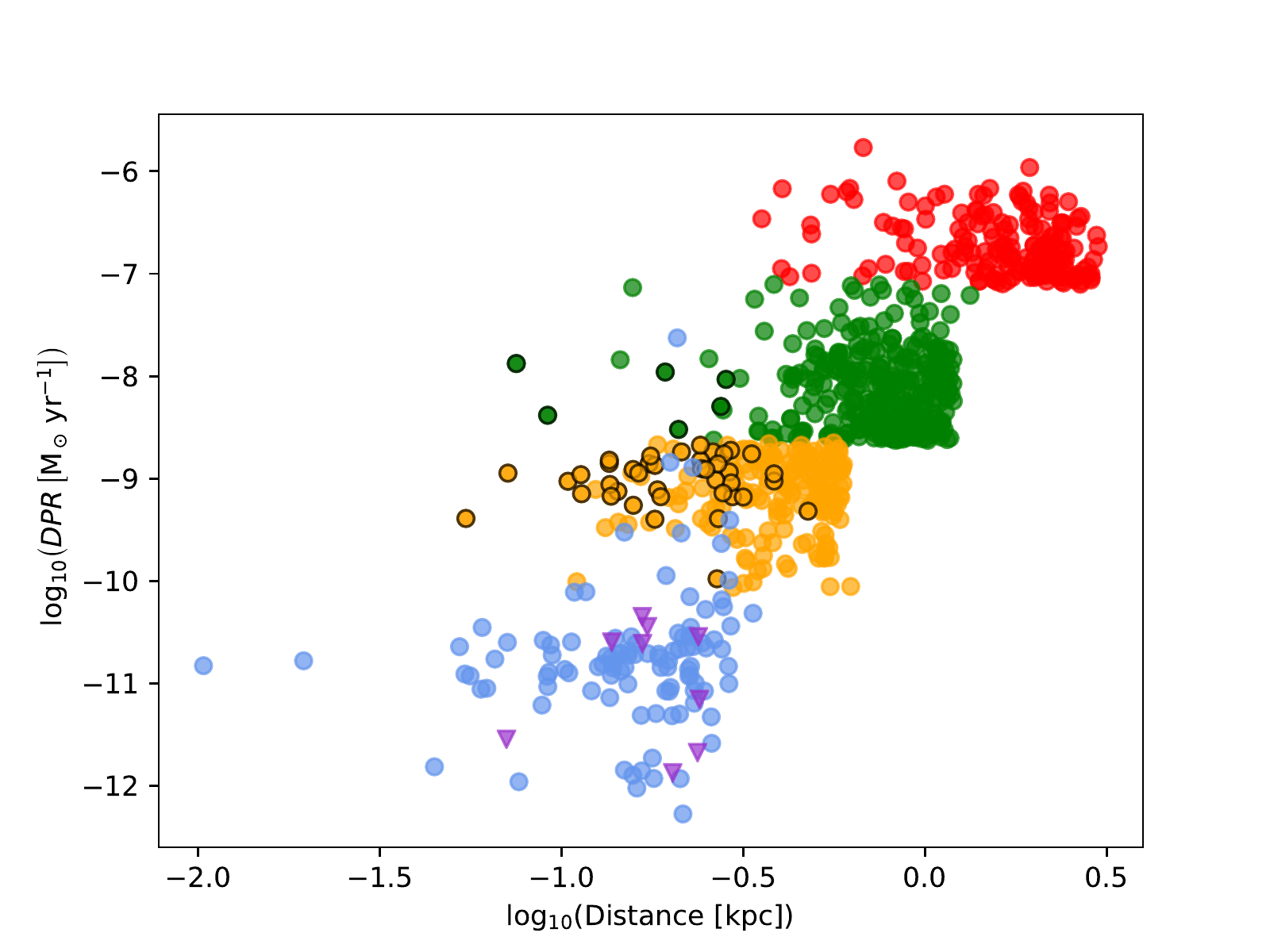}
    \caption{The NESS sample in distance -- dust-production rate space. The colours correspond to the different survey tiers (Purple: tier 0 (DPRs are upper limits); Blue: tier 1; Orange: tier 2; Green: tier 3; Red: tier 4); points with black outlines indicate sources selected for mapping. 
    }
    \label{fig:sampledpr}
\end{figure}

\begin{figure}
    \centering
    \includegraphics[width=0.475\textwidth, clip=true, trim=0.65cm 2.5cm 1.6cm 2.65cm]{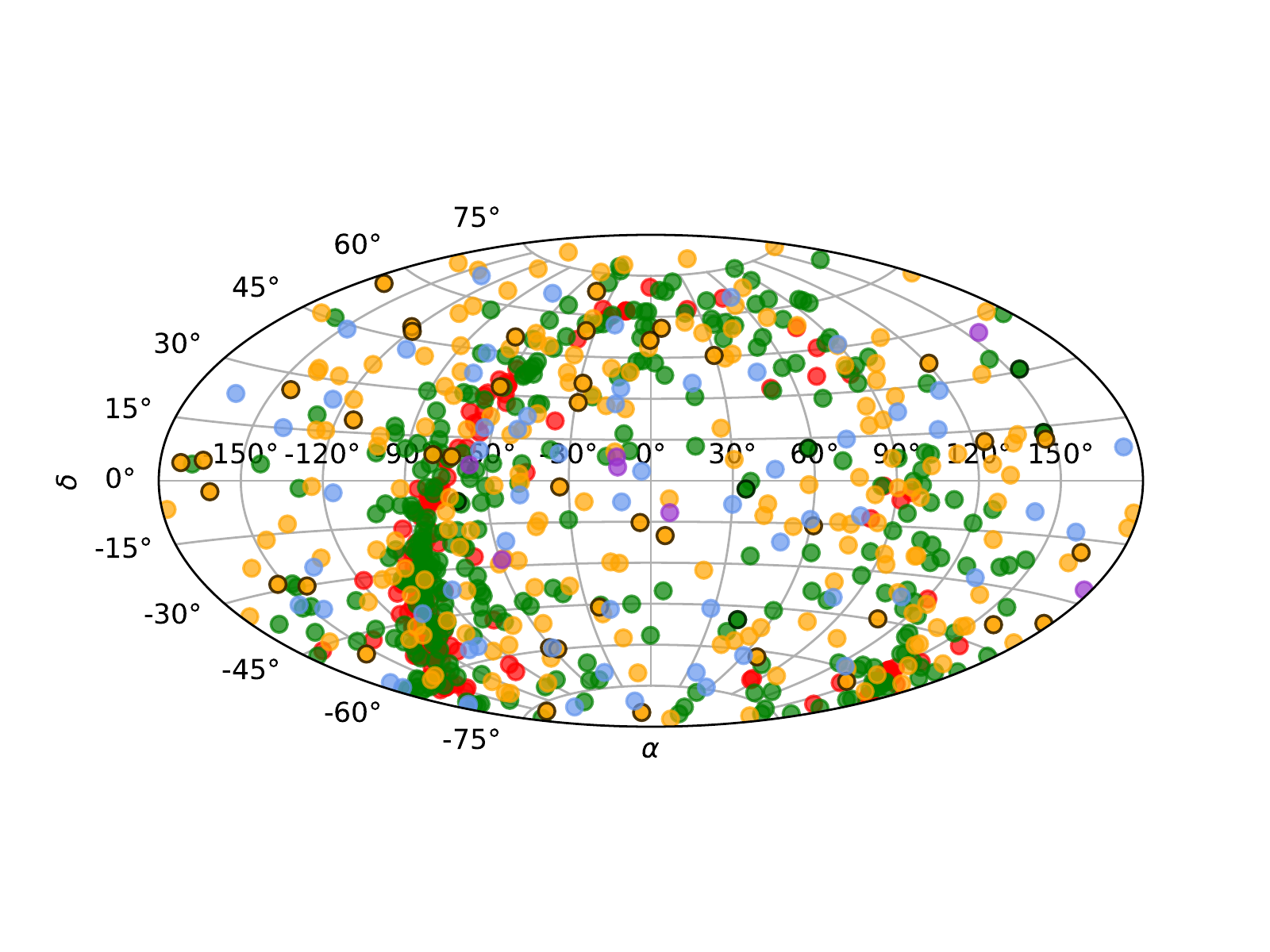}
    \caption{Sky distribution of NESS sources, using the same colour scheme as Fig.~\ref{fig:sampledpr}.}
    \label{fig:samplesky}
\end{figure}

\begin{table*}
    \centering
    \caption{Table of NESS sources showing first 10 sources to demonstrate the format of the data. A larger table is available online. The sources of the distances are indicated with M = maser parallax, H = \emph{Hipparcos} parallax, G = \emph{Gaia}/TGAS parallax and L = luminosity distance. The chemical types are derived from spectroscopy as described in Sect.~\ref{sec:samp}.
    }
    \begin{tabular}{lllllcccc@{\ \ }c@{\ \ }c@{\  \ }c}
    \hline    IRAS PSC \# & \multicolumn{1}{c}{\sc simbad} &\multicolumn{1}{c}{2MASS} & \multicolumn{1}{c}{$\alpha$} & \multicolumn{1}{c}{$\delta$} & d &Dist.& $\dot{M}_{\rm dust}$&Chem& NESS & Mapping \\ 
     &\multicolumn{1}{c}{ID}&\multicolumn{1}{c}{ID}&\multicolumn{1}{c}{(J2000.0)}&\multicolumn{1}{c}{(J2000.0)}& (kpc)&type &(M$_{\odot}$\,yr$^{-1})$ &type& tier&\\\hline \hline
     IRAS 00042+4248 & V* KU And & 00065274+4305021 & 00 06 52.75 & +43 05 02.25 & 0.54 & L & 1.7 $\times$ 10$^{-8}$ & O & 3 & 0 \\
     IRAS 00084-1851 & V* AC Cet & 00105796-1834224 &00 10 57.95  &-18 34 22.55&0.29&H&2.1$\times$ 10$^{-10}$&C&1&	0 \\		
     IRAS 00121-1912 & V* AE Cet & 00143841-1855583 & 00 14 38.42 & -18 55 58.31 & 0.14 & H & 7.5 $\times$ 10$^{-12}$ &  & 1 & 0 \\
     IRAS 00192-2020 & V* T Cet & 00214626-2003291 & 00 21 46.27 & -20 03 28.88 & 0.27 & H & 5.2 $\times$ 10$^{-10}$ & O & 2 & 1 \\
     IRAS 00193-4033 & V* BE Phe & 00214742-4017155 &00 21 47.42 &-40 17 15.47 & 0.92 & L & 7.1 $\times$ 10$^{-8}$ & O & 3 & 0 \\
     IRAS 00205+5530 & V* T Cas & 00231427+5547332 & 00 23 14.27 & +55 47 33.21 & 0.29 & L & 1.5 $\times$ 10$^{-9}$ & O & 2 & 1 \\
     IRAS 00213+3817 & V* R And & 00240197+3834373 & 00 24 01.95 & +38 34 37.35 & 0.43 & L & 4.9 $\times$ 10$^{-9}$ & S & 3 & 0 \\
     IRAS 00245-0652 & V* UY Cet & 00270644-0636168 &00 27 06.45 & -06 36 16.87 & 0.45 & L & 1.2 $\times$ 10$^{-9}$ & O & 2 & 0 \\
     IRAS 00247+6922 & V* V668 Cas & 00274110+6938515 &00 27 41.13 &+69 38 51.61 & 0.94 & L & 1.2 $\times$ 10$^{-8}$ & C & 3 & 0 \\
     IRAS 00254-1156 & V* AG Cet & 00280053-1139318 &00 28 00.55 &-11 39 31.68 & 0.24 & H & 4.2 $\times$ 10$^{-13}$ &  & 0 & 0 \\
\hline
    \end{tabular}
    \label{tab:sample}
\end{table*}

Using the distances and dust-derived $\dot{M}$ derived above, we define five sub-samples of sources to be observed (Fig.~\ref{fig:sampledpr}). 
These samples are defined in terms of increasing distance and $\dot{M}_{\rm dust}$.
\begin{itemize}
    \item Tier 0  (or ``very low'' DPR sources) is a special addition of ten sources with $L \geq 1600$\,L$_\odot$, $d<250$\,pc and $\delta \geq -30\deg$, without limit on $\dot{M}_{\rm dust}$, drawn from \citet{McDonald12} and \citet{McDonald17}. This tier explores mass loss from bright red giant branch (RGB) and AGB stars not producing dust. Nine more sources at $\delta \leq -30\deg$ meet these criteria, but have not yet been scheduled for observation by NESS. One (SX Pav) has been published already \citep{McDonald18}.
    \item Tier 1 ( ``low''; 105 sources) includes all sources with $\dot{M}_{\rm dust}<10^{-10}$ M$_\odot$ yr$^{-1}$ at $d<300$\,pc, and samples the AGB stars with the lowest $\dot{M}_{\rm dust}$. Sources were only included if there was a 3$\sigma$ dust excess in the GRAMS models, or if the source has an infrared excess in \citet{McDonald12} or \citet{McDonald17}.
    \item Tier 2 ( ``intermediate''; 222 sources) contains all sources with $10^{-10} \leq \dot{M}_{\rm dust} < 3\times10^{-9}$ M$_\odot$ yr$^{-1}$ and $d<600$\,pc, excluding the Galactic plane ($\left|b\right|<1.5$) for $d>400\,pc$;
    \item Tier 3 ( ``high''; 324 sources) consists of all sources with $3\times10^{-9} \leq \dot{M}_{\rm dust} < 10^{-7}$ M$_\odot$ yr$^{-1}$ and $d<1200$\,pc, excluding the Galactic plane for $d>800$\,pc; and
    \item Tier 4 ( ``extreme''; 182 sources) comprises all sources with $\dot{M}_{\rm dust} \geq 10^{-7}$ M$_\odot$ yr$^{-1}$ and $d<3000$\,pc, excluding the Galactic plane for $d>2000\,pc$.
\end{itemize}
These tier limits result in each of Tiers 1--4 containing enough objects ($>$100) to adequately establish the typical range of properties of AGB winds within that tier, while the limit of $L \geq 1600$ L$_\odot$ approximates the typical luminosity at which dust-production begins for the lowest-mass stars \citep[e.g.][]{McDonald11,McDonald11b,McDonald14}. 
Sources in Tier 4 are poorly classified in the literature, so are more likely to be contaminants. The NESS survey aims to improve on these objects' classifications.
Since the original selection, updated distances for some sources have shifted their locations in Fig.~\ref{fig:sampledpr}.

Of the entire sample of 852 sources,
Table~\ref{tab:sample} gives the first ten entries, sorted by right ascension, to demonstrate the information included and its format. For each source, we report the distance used, the source of this distance,
the GRAMS best-fit $\dot{M}$, and classifications based on \emph{IRAS} LRS or \emph{ISO} SWS spectra when available. The S-star classifications are from \citet{SloanPrice1998}, \citet{Yangetal2007}, and \citet{Honyetal2009}, with the \citet{Honyetal2009} classifications taking precedence.
The full interactive table, available at \url{https://evolvedstars.space}, represents the initial state of the NESS catalogue.
Future data releases will attach the stellar parameters assumed and NESS data products to each source.
Figure \ref{fig:samplesky} illustrates the sky distribution.

\begin{figure}
    \centering
    \includegraphics[width=0.475\textwidth]{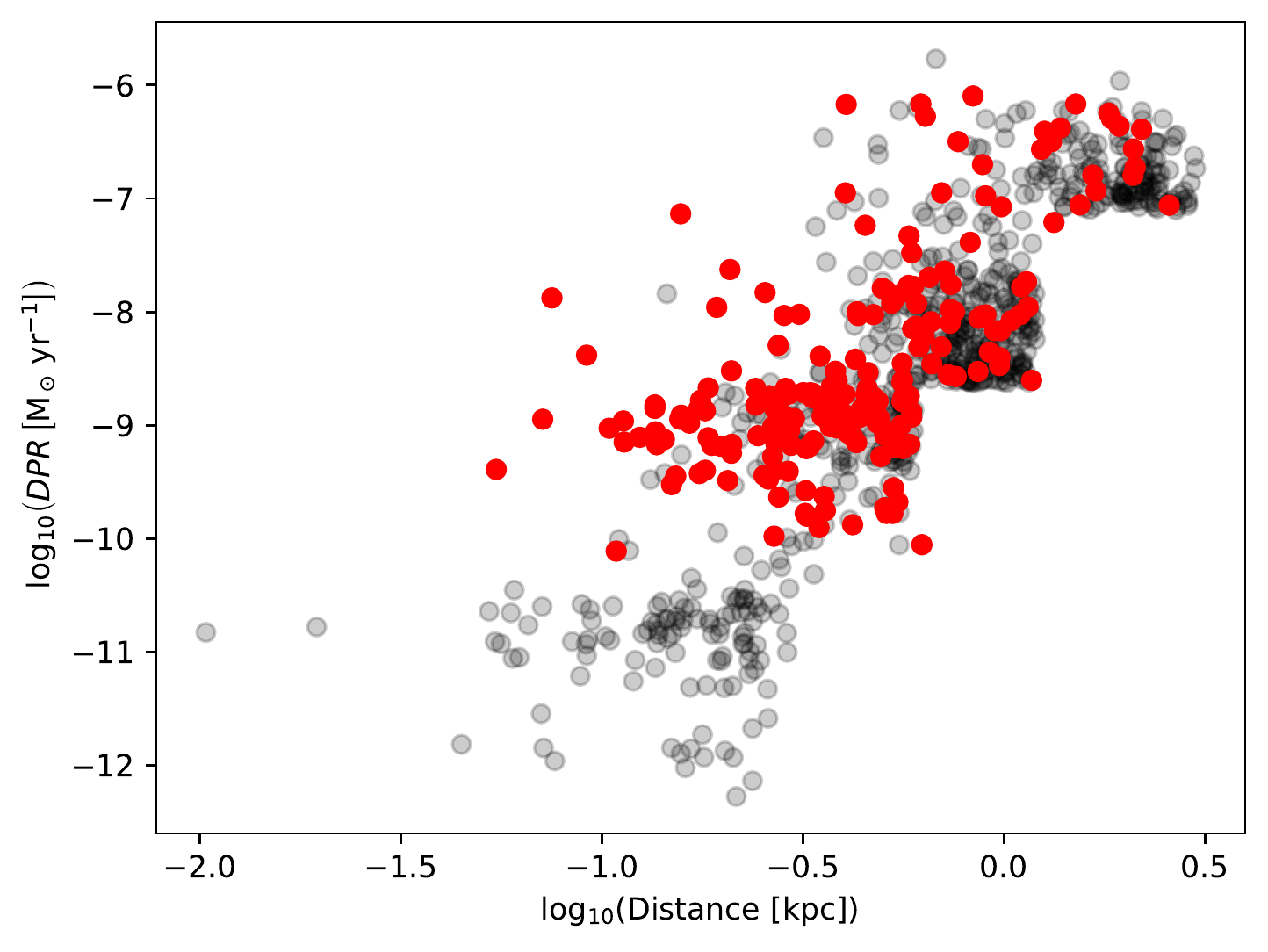}
    \caption{The NESS sample as in Figure~\ref{fig:sampledpr} (shown in grey) highlighting sources with detections in CO(2--1) or (3--2) from previous surveys (red).}
    \label{fig:nessvslit}
\end{figure}

 We find that the CO studies discussed above in Sect.~\ref{sec:costudies} detect only 204 of the 852 NESS targets (less than 25\%) in either or both of the CO $J$=2--1 and $J$=3--2 transitions. Figure \ref{fig:nessvslit} shows that the NESS survey will extend these observations to the sources with least dust content, which are almost completely missing in previous studies. Furthermore, NESS will provide a four-to-five-fold improvement in the statistics at DPR~$\geq 5\times10^{-9}$\,M$_\odot$ yr$^{-1}$ and at distances greater than 600\,pc.

\subsubsection{Mapping sample}\label{subsec:mapping}
In addition, 46 of the nearer sources in the staring sample with significant dust production were selected for mapping on arcminute scales (black circles in Fig.~\ref{fig:sampledpr}). 
These comprise stars outside the Galactic Plane ($|b| \geq 1.5$ deg) with $\dot{M}_{\rm dust} > 5\times10^{-10}$ M$_\odot$\,yr$^{-1}$ 
and $d < 340$\,pc, which are bright and whose CO envelopes have a large predicted angular size ($\gtrsim 20\arcsec$, based on \citealt{Mamon88}). Although the GRAMS best-fit prediction for U Ant was lower than the cutoff, it is known to have an extended envelope in the FIR \citep[e.g.][]{Kerschbaum2010}, such that GRAMS underestimates the DPR. It has been published separately as \citet{Dharmawardena2019}.
Seven sources satisfying these criteria were not selected into the mapping sample; they will be folded in as part of future proposals to complete this set.

\subsubsection{Archival sub-mm data}

Many stars in our sample have existing JCMT or APEX observations of the (2--1) and (3--2) transitions of $^{12}$CO and $^{13}$CO, or existing JCMT continuum data from SCUBA-2, and we use these archival data rather than re-observing these sources. Data from JCMT/SCUBA and APEX (both LABOCA and SABOCA) at the same wavelengths are not included because the fields of view are much smaller.

\subsubsection{\emph{Gaia} EDR3}\label{sec:gaiaEDR3}

\begin{figure}
    \centering
    \includegraphics[width=0.475\textwidth]{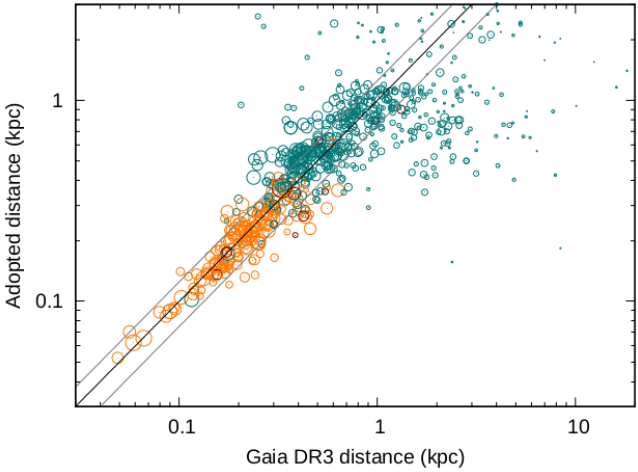}
    \caption{Comparison of adopted distances and distances from \emph{Gaia} EDR3 parallaxes. Colours represent adopted distances from masers (dark red), parallax from \emph{Hipparcos} or TGAS (orange), and luminosity distances (cyan). Diagonal lines represent parity and $\pm$25 per cent, i.e., the anticipated uncertainty of our luminosity distances. Point area is proportional to precision in \emph{Gaia} DR3 parallax ($\varpi/\sigma_\varpi$, i.e. bigger points have better \emph{Gaia} precision).
    }
    \label{fig:gaiaedr3}
\end{figure}

\emph{Gaia} Early Data Release 3 (EDR3) occurred during the refereeing process of this paper. Its 34-month timespan (comparable to that of individual \emph{Hipparcos} stars) significantly reduces astrometric noise compared to DR2, improving parallaxes for nearby evolved stars. While this improves the distances of many of our sources, we here retain the distances used to define the membership of our catalogue tiers for self-consistency, and defer decisions on individual distances to a dedicated catalogue paper (McDonald et al., in prep.).

Meanwhile, we compare the \emph{Gaia} EDR3 distances to those used to select our sources. To identify \emph{Gaia} EDR3 cross-matches to our \emph{IRAS} sources, we selected \emph{Gaia} sources within 3$^{\prime\prime}$ of the {\sc simbad} co-ordinates, provided they had red \emph{Gaia} colours ($B_P - R_P > 1$ mag; 771 sources cross-matched), and compute distances for these by na\"ively inverting the parallax (see comments below).

As \emph{Gaia} EDR3 is largely independent of our source data (only a few sources were selected using TGAS parallaxes), this comparison tests the accuracy of our distance-selection methods. Figure \ref{fig:gaiaedr3} shows a good correlation overall. The scatter comes from the combined errors in the \emph{Gaia} EDR3 and adopted distance data. We assume they are uncorrelated, thus
\begin{equation}
    \sigma\left[{\rm Gaia} / {\rm adopted}\right] \approx \sqrt{\left(\frac{\sigma_{\rm Gaia}}{d_{\rm Gaia}}\right)^2 + \left(\frac{\sigma_{\rm adopted}}{d_{\rm adopted}}\right)^2} .
    \label{eq:sigmaplx}
\end{equation}

A long tail of sources towards large \emph{Gaia} distances, caused by parallaxes with large fractional errors, skews the sample from a normal distribution. Some of these stars could be RSGs. However, this long tail continues into negative parallaxes for 31 sources (3 per cent), four of which are statistically significant at the 4--6$\sigma$ level. A further 47 sources have a \emph{Gaia} cross-match with no determined parallax. Many sources in this long tail have astrometric noise exceeding the parallax itself, showing photo-centric motion is still a significant issue for AGB stars in \emph{Gaia} EDR3. 

A few sources also scatter to lower \emph{Gaia} distances than expected, mostly in Tier 2. Some may be RGB stars, which are not contained in the prior luminosity function from the LMC. However, their inclusion requires detectable infrared excess, which is generally not observed in RGB stars \citep[e.g.][]{McDonald11c}. As the distribution is markedly non-Gaussian at both under- and over-estimated distances, we investigate the central 68th centile of the distribution of the three $\sigma$ terms in Equation \ref{eq:sigmaplx}, rather than their standard deviation. This avoids detailed treatment of stars with poor, zero and negative parallaxes, as these are rare or non-existent within the central 68th centile.

Where the \emph{Hipparcos} parallax is the adopted distance measure, $\sigma$[\emph{Gaia} / \emph{Hipparcos}] = 0.16. These are typically close stars (median \emph{Hipparcos} distance = 209 pc), with low astrometric noise, and the quadrature addition of fractional parallax errors (Equation \ref{eq:sigmaplx}) is only slightly more than the value of 0.12 expected from the quoted astrometric uncertainties of the \emph{Hipparcos} and \emph{Gaia} surveys.

Beyond $\sim$400 pc, imprecise historical parallaxes become replaced by luminosity distances. In tier 2 (including four sources with luminosity distances from tier 1), $\sigma$[Gaia / adopted] = 0.36, on average, for a median adopted distance of 477 pc. Similar numbers for tiers 3 and 4 are 0.45 (849 pc) and 0.84 (2121 pc). If we first assume $\sigma_{\rm adopted}/d_{\rm adopted} = 0.25$, then we can approximate the true uncertainty in \emph{Gaia} EDR3 for tiers 2, 3 and 4 as $\sigma_{\rm Gaia}/d_{\rm Gaia}$ = 0.26, 0.38 and 0.80, respectively.

While $\sigma_{\rm adopted,lum}$ should be largely invariant with true distance, $\sigma_{\rm Gaia}$ should increase linearly with true distance. We can enforce this by setting $\sigma_{\rm adopted,lum} = 0.316$. However, the extremity of our sources (thus their astrometric noise) does increase with distance, while their brightness decreases as they become more optically obscured, and they become more crowded as they concentrate closer to the Galactic plane (Section \ref{sec:galactocentric}). Consequently, $\sigma_{\rm Gaia}$ will increase with distance, $\sigma_{\rm adopted,lum} = 0.316$ can be treated as an upper limit, and we estimate that our adopted value ($\sigma_{\rm adopted,lum} = 0.25$) is approximately correct.

We can also assume that the ratio of \emph{Gaia} EDR3 distances to luminosity distances should have a median of unity within each tier\footnote{We have not performed the parallax bias correction of \citet{Lindegren20}. { Of the 771 \emph{Gaia} EDR3 cross-matches, only 183 have solution types or colours within the range where these corrections are valid. For these remaining sources, the correction requires} an accurate and invariant colour and magnitude, hence a detailed treatment of the epoch photometry, to be precise. The corrections for these 183 sources based on their given colours and magnitudes are fairly small ($<$52 $\mu$as and $<$9.6 per cent in 95 per cent of cases) and less than the parallax error in all but five cases. Consequently, we ignore the correction.}. Where we have adopted maser distances, the median $d_{\rm adopted,maser}/d_{\rm Gaia} = 0.80$ with a standard error of 0.17. Where we have adopted parallaxes, $d_{\rm adopted,plx}/d_{\rm Gaia} = 0.97 \pm 0.03$. Finally, for our luminosity-based distances in tiers 2, 3 and 4, respectively, $d_{\rm adopted,lum}/d_{\rm Gaia}$ = 0.98 $\pm$ 0.07, 0.85 $\pm$ 0.07, 1.02 $\pm$ 0.26. While this generates a 2.0$\sigma$ outlier for tier 3, the sample overall is in statistical agreement with $d_{\rm adopted,lum}/d_{\rm Gaia}$ = 1. Hence we consider there to be no discernable systematic offset between our adopted distances and the true distance, despite the varying and competing effects of Malmquist and Lutz--Kelker biases in the Galactic and LMC samples, and the differing properties of the two galaxies.

\subsubsection{Completeness}\label{sec:completeness}

Since a substantial part of the infrared emission of high-mass-loss-rate AGB stars is from the dust, the completeness of \emph{IRAS} to \emph{AGB} stars at a given distance is a function of dust-production rate. The 2.5 Jy cut (Section \ref{sec:objects}) should retain all AGB stars brighter than the tip of the RGB out to $\sim$500 pc, hence we can consider our sample of AGB stars complete to these distances for all tiers. It should further retain the brightest objects and those with any measurable mass loss out to much greater distances, and we have checked that the distance distribution of each tier beyond $\sim$500 pc approximates the $d^2$ distribution expected for the Galactic disc, allowing for stochastic variance and the missing sample from $\left|b\right|<1.5$. Since AGB stars are intrinsically bright in the infrared, we do not expect significant source confusion either at small distances, or at larger distances when $\left|b\right|<1.5$. 

If the 60 $\mu$m flux were the only constraint, our limiting DPR would be proportional to the square of the distance. Tiers 4 and 3 would then be more than 95\% complete, but Tier 2 would be incomplete at large distances (missing some sources with lower DPRs). Tier 1 would remain unaffected thanks to the inclusion of nearby sources from \citet{McDonald12, McDonald17}.

However, the 60 $\mu$m flux is not linearly dependent on the DPR because, below a DPR of roughly $0.3 - 3 \times 10^{-9}$ M$_\odot$ yr$^{-1}$, the photosphere becomes the dominant contributor to the far-IR emission. Since our sample is complete to naked (zero-DPR) stars at the tip of the RGB out to at least 500 pc, under the worst-case scenario the outer 100 pc of Tier 2 would become incomplete. However, since the minimum DPR in Tier 2 is non-zero, the combined emission from dust and photosphere are expected to render our sample in Tier 2 complete as well. It is worth noting that changes in gas-to-dust ratio impact these calculations: our sample becomes more complete as the ratio increases.

\section{Observing strategy}\label{sec:obs}

This section describes the observing strategy for heterodyne observations of the CO(2--1) and (3--2) lines and bolometer observations of the sub-mm continuum of targets with $\delta > -40^{\circ}$, which are being observed with the JCMT\footnote{Program IDs: M16XP001, M17AP027, M17BL002, M19BP035, M20AL014} (500 out of 852 sources), with the RxA3m, `\=U`\=u\footnote{`\=U`\=u replaced RxA3m in September 2019, after RxA3m was decommissioned in 2018} \citep{Mizuno2020}, HARP \citep{Buckle2009}, `\=Aweoweo\footnote{`\=Aweoweo is a new 345\,GHz receiver currently undergoing commissioning} \citep{Mizuno2020} and SCUBA-2 \citep{Holland2013} instruments, respectively, avoiding sources in the Galactic plane ($\left|b\right| \leq 1.5^\circ$).
These will be complemented with observations from APEX\footnote{Program IDs: O-0101.F-9308A,B-2018, E-0101.D-0624A,B-2018 and M-0105.F-9534A,C-2020} for southern sources  (291 out of 852 sources), 
the strategy for which will be described in Wallstr\"om et al., in prep.
Similarly, sources with $\left|b\right| \leq 1.5^\circ$ have been assigned for future interferometric observation to mitigate confusion from interstellar lines (128 out of 852 sources). These are mostly in the Galactic plane at great distance, i.e., belong mostly to Tier 4. A summary of our strategy is given in Table~\ref{tab:obsstrat}.
 All sensitivities are given as the expected RMS\ noise level, i.e., the $1\sigma$ sensitivity.

\begin{table*}
    \caption{Summary of observing strategy}
    \label{tab:obsstrat}
    \centering
    \begin{tabular}{cccc}
        \hline
        \multirow{2}{*}{Subsample} & \multicolumn{3}{c}{Stategy} \\
        &Continuum&345\,GHz&230\,GHz\\\hline\hline
        \multirow{2}{*}{Tier 0} &\multirow{2}{*}{--}&\multirow{2}{*}{--} & CO: 0.003\,K\,T$_{\rm A}^{\ast}$\\
         &&&$^{13}$CO$^{a}$: 0.003\,K\,T$_{\rm A}^{\ast}$\\
        \multirow{2}{*}{Tier 1} &850: 3\,mJy\,beam$^{-1}$&\multirow{2}{*}{--}
        & CO: 0.01\,K\,T$_{\rm A}^{\ast}$\\
         &450: no constraint&&$^{13}$CO$^{a}$: 0.01\,K\,T$_{\rm A}^{\ast}$\\
        \multirow{2}{*}{Tier 2 -- 4} &850: 3\,mJy\,beam$^{-1}$ &CO: 0.025\,K\,T$_{\rm A}^{\ast}$&CO: 0.01\,K\,T$_{\rm A}^{\ast}$ \\
         &450: no constraint&$^{13}$CO$^{a}$: if CO detected $\geq$0.3\,K\,T$_{\rm A}^{\ast}$&$^{13}$CO$^{a}$: 0.01\,K\,T$_{\rm A}^{\ast}$\\
        \multirow{2}{*}{Mapping} &850: 1.5\,mJy\,beam$^{-1}$ &CO: 0.07\,K\,T$_{\rm A}^{\ast}$& CO: 0.07\,K\,T$_{\rm A}^{\ast}$\\
        &450: no constraint&No $^{13}$CO$^{a}$&No $^{13}$CO$^{a}$ \\ \hline
        \multicolumn{4}{l}{$^a$: $^{13}$CO data may be observed ``for free'' in the lower sideband of observations with APEX or \uu.} 
        \\ \hline
    \end{tabular}
    
\end{table*}

\subsection{Staring}
Staring sources are observed in different setups, depending on their $\dot{M}_{\rm dust}$.
Tier 0 sources, nominally dust-free and with low luminosities, are observed only with `\=U`\=u. We exploit the substantial improvement in sensitivity compared to older receivers to target a sensitivity of 0.003\,K\,T${\rm _A^\ast}$ at a resolution of 1~km\,s$^{-1}$ in the CO(2--1) line. While tier 0 sources are not observed in continuum because of their low dust content, these gas tracers can then be compared with mid-infrared probes of the dust component of the outflow.
Tier 1 sources are observed only in CO (2--1) with RxA3m or `\=U`\=u, obtaining a single, deep spectrum with a target noise level 0.01 K $T{\rm _A^\ast}$ at 1~km\,s$^{-1}$ under weather conditions where the opacity at 225\,GHz as measured by the JCMT water-vapour radiometer $\tau_{225} \geq 0.2$. The sensitivity is chosen to achieve a 3$\sigma$ detection of the observed CO flux of EU Del\footnote{EU Del was chosen as the lowest-luminosity and lowest-mass-loss-rate evolved star that had been successfully detected at the foundation of the survey. This ignores the lower mass-loss-rate star VY Leo \citep{Groenewegen14}, which we could not expect to detect in a reasonable amount of time.} at a resolution of 1~km\,s$^{-1}$ at a distance of 300 pc \citep{McDonald18}. Sources detected in CO (2--1) are then observed in CO (3--2). Tier 1 sources are also observed in continuum. The continuum observations consist of a single 31-minute scan using the CV Daisy scan-pattern, which performs a nearly circular scan at a constant scan speed of 155\,\arcsec\,s$^{-1}$ while ensuring that the target remains on the detector arrays at all times during the observation. 
This mapping mode results in a map with a diameter of $\sim$\,15\arcmin; with sensitivity roughly uniform in the central 3\arcmin\ of the map, and declining towards the outskirts of the map.
Previous studies have shown that this is sufficiently large to observe nearby AGB stars \citep[e.g.][]{Dharmawardena2018}.
We target a sensitivity of 3~mJy\,beam$^{-1}$ at 850\,$\mu$m in the central 3\arcmin\ region of the map for $\tau_{225} \leq 0.08$.

All other sources are observed in CO (2--1), CO(3--2), and also 850 and 450\,$\mu$m continuum. 
Our original strategy was for sources which are brighter than 0.3\,K in the $^{12}$CO lines to also be observed in matching $^{13}$CO transitions. 
However, the introduction of `\=U`\=u at the JCMT means that effectively all sources have both CO isotopologue\,($2-1$) lines observed simultaneously and to the same depth, thanks to the dual sideband observations; sources observed at 345\,GHz with APEX also have simultaneous $^{13}$CO (3--2) observations, but our original strategy remains for observing CO (3--2) with the JCMT.
The CO line observations aim to achieve a noise level of $\sim$0.01\,K\,T${\rm _A^\ast}$ at 1 km\,s$^{-1}$ resolution, under weather conditions of $\tau_{225} \geq 0.2$ for RxA3m or `\=U`\=u  and $\tau_{225} \leq 0.2$ for HARP. 
The continuum observations are identical to those in tier 1.

\subsection{Mapping}

As the mapping sources have CO lines extending over radii $\gtrsim 20\arcsec$ from the star (Sect.~\ref{sec:samp}), these sources are observed in raster (RxA3m or `\=U`\=u) or jiggle (HARP) modes to produce a $2^\prime \times 2^\prime$ map of the line emission. 
We aim for a noise level of $\sim$0.07\,K\,T${\rm _A^\ast}$ at 1~km\,s$^{-1}$ resolution over the map for each source under weather conditions of $\tau_{225} \leq 0.2$ for RxA3m and `\=U`\=u and $\tau_{225} \leq 0.12$ for HARP. This noise level was chosen to achieve dynamic range in excess of 100 for the brightest sources in the sample.

The continuum observations employ the strategy demonstrated by \citet{Dharmawardena2018}.
For each source, four 31-minute scans were used, with the southernmost sources ($\delta \leq -30^\circ$) observed in better weather ($\tau_{225} \leq 0.05$ rather than $0.08$), resulting in an estimated sensitivity of $\sim$ 1.5~mJy\,beam$^{-1}$ at 850\,$\mu$m\footnote{The sensitivity at 450\,$\mu$m has a much stronger dependence on the atmospheric conditions. We therefore do not place strong constraints on the required depth at 450\,$\mu$m.}.
The SCUBA-2 850\,$\mu$m filter includes the CO(3--2) line, which is typically the strongest line in this band pass for all known chemical types of AGB stars, and may contribute to the observed flux \citep[][see also below]{Drabek2012}. 
While the 450\,\micron\ band also includes the CO(6--5) line, the continuum is typically much brighter, while the line is typically of similar or lower strength than CO(3--2) \citep{Kemper03, DeBeck10}, making the typical line-to-continuum ratio smaller.

\subsection{Survey Progress}\label{sec:prog}

As of December 2020, NESS has taken 772 hours of observations: 666 at the JCMT across two large programs and three PI programs, and 106 at APEX across three observing programs, representing 51.1 per cent (time) completion.
NESS has observed 252 sources in continuum and 746 sources in at least one CO line. Detection rates among this sample are $\sim$70 per cent in both $^{12}$CO lines and $\sim$30 per cent in both $^{13}$CO lines. The new `\=U`\=u and `\=Aweoweo receivers are more sensitive, and many sources have not yet been observed to their full requested depth, hence we expect these rates to increase substantially over the full survey.

\subsection{Data reduction}\label{sec:red}
Data-reduction methods and scripts for our observations will be presented alongside survey measurements in forthcoming papers:
Wallstr\"om et al. (in prep.) will present the heterodyne data, while Dharmawardena et al. (in prep.) will present SCUBA-2 observations. Here, we present an initial reduction of the SCUBA-2 data. 

\begin{figure}
    \centering
    \includegraphics[width=0.475\textwidth, clip=true, trim=0.5cm 0.25cm 1.5cm 1.45cm]{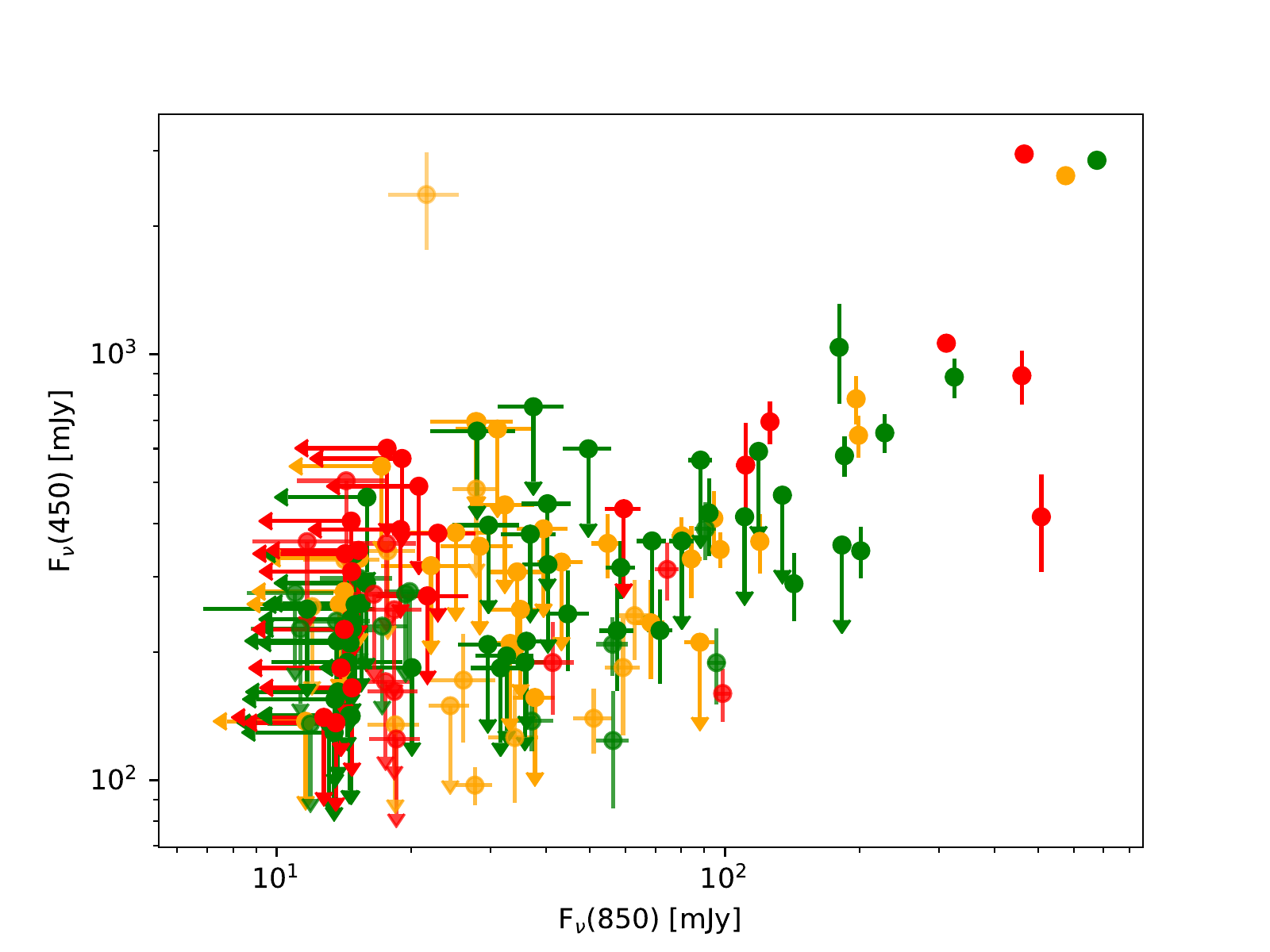}
    \caption{SCUBA-2 fluxes for the subset of NESS sources observed to date, using the same colour code as Fig.~\ref{fig:sampledpr}. 
    }
    \label{fig:scuba2fluxes}
\end{figure}

For this paper, we adopt a simple data-reduction strategy to measure fluxes for the point-source components of NESS sources at 450\,and 850\,\micron, ignoring extended emission. 
Using the default masking and filtering parameters in Starlink/ORAC-DR, the SCUBA-2 pipeline \citep{Chapin2013,Jenness2015}, we reduced all SCUBA-2 observations taken as part of the NESS programme before 2019 May 27, excluding the sources selected for mapping, giving a143 sources with observations with angular resolutions of $\sim 7.9\arcsec$ and $\sim 14\arcsec$ at 450 and 850\,\micron, respectively. 
The images are calibrated in mJy\,beam$^{-1}$, and the flux in the brightest pixel extracted as an estimate of the point-source flux. 
If no source was detected at the 4$\sigma$ level, the reduction was repeated using matched filtering, which optimises the sensitivity to point-like sources.
Since matched filtering systematically underestimates fluxes, we multiply the flux of sources detected using this method by 1.1 \citep{Geach2013MNRAS.432...53G, Smail2014ApJ...782...19S}. 
We note that no subtraction of CO line emission has been performed here, suggesting that the fluxes at 850\,\micron\ may be overestimated by 10 -- 20 per cent \citep[e.g.][]{Drabek2012, Dharmawardena2019}.
The fluxes or 3-sigma upper limits are shown in Fig.~\ref{fig:scuba2fluxes} and the first  10 sources are shown in Tab.~\ref{tab:flux}.
The 450\ \micron\ flux is typically a factor of a few greater than at 850 \micron, a trend that the upper limits are not in conflict with. 
There are two objects that deviate from this trend; one planetary nebula whose 850\ \micron\ emission is comparable to its 450\ \micron\ emission, and one AGB star with an anomalously high 450\ \micron\ flux and a high space motion. These sources and the reason for their unusual fluxes are discussed in Sect.~\ref{sec:cold}.

\begin{table}
\caption{Continuum fluxes and spectral indices {($\alpha$, see Sect.~\ref{sec:cold})} from the initial reduction of the SCUBA-2 data presented here. Limits correspond to 3 times the RMS of the map in mJy/beam.  The first 10 sources are shown here, the full sample is available online.}
    \centering
    \begin{tabular}{lccc}
\hline IRAS PSC \# & F$_{450}$ & F$_{850}$ & $\alpha$ \\ 
 & mJy & mJy & \\\hline\hline 
IRAS 00042+4248&$<364.2$&$68.8\pm5.5$&$>-3.9$ \\
IRAS 00192-2020&$<251.6$&$35.0\pm5.4$&$>-4.5$ \\
IRAS 01159+7220&$<415.2$&$110.7\pm8.9$&$>-3.3$ \\
IRAS 02316+6455&$<320.5$&$40.3\pm5.0$&$>-4.7$ \\
IRAS 02351-2711&$<599.3$&$49.6\pm6.1$&$>-5.5$ \\
IRAS 03149+3244&$<225.8$&$<14.2$&$>-12.6$ \\
IRAS 03170+3150&$<212.1$&$<13.6$&$>-12.7$ \\
IRAS 03206+6521&$<363.0$&$11.7\pm2.9$&$>-7.3$ \\
IRAS 03229+4721&$<356.2$&$182.5\pm14.6$&$>-2.1$ \\
IRAS 04166+4056&$<189.0$&$14.4\pm4.6$&$>-5.8$ \\
\hline
    \end{tabular}
    
    \label{tab:flux}
\end{table}

As these reductions are preliminary, we aim to be conservative about the flux measurements and their uncertainties: by only measuring the point-like component of the flux, we minimise the influence of the artefacts that commonly affect SCUBA-2 observations, known as negative bowling and blooming, where the emission is filtered too aggressively or where sky emission is mistakenly identified as astronomical, respectively.
These problems particularly affect the 450\,$\mu$m data, as the atmosphere fluctuates more than at 850\,$\mu$m; hence, we validate our results by comparing the SCUBA-2 fluxes to \emph{Herschel Space Telescope} Spectral and Photometric Imaging Receiver (\emph{Herschel}/SPIRE) fluxes where available.

\begin{figure}
    \centering
    \includegraphics[width=0.475\textwidth, clip=true, trim=0.25cm 0.25cm 1.cm 1.cm]{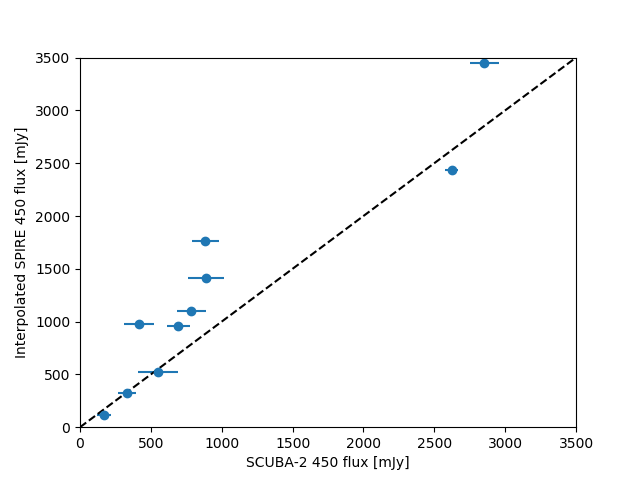}
    \caption{Comparison of 450 fluxes observed with SCUBA-2 to those estimated by interpolating between SPIRE fluxes. The dashed line indicates the 1:1 relation.}
    \label{fig:s2spire}
\end{figure}

Of the 143 sources for which SCUBA-2 fluxes are presented, we find ten objects that are detected in both the SPIRE 350 and 500\,\micron\ bands.
We interpolate between the two SPIRE fluxes to estimate a 450\,\micron\ flux as
\begin{equation}
    F_{450} = F_{350} \times \left(450/350\right)^{\alpha_{\rm SPIRE}}
\end{equation}
where $$\alpha_{\rm SPIRE} = \frac{\log\ (F_{350}/F_{500})}{\log(350/500)}.$$
For these ten objects, we compare the observed and estimated 450\,\micron\ fluxes in Figure~\ref{fig:s2spire}.
In this figure, we can clearly see that all sources lie reasonably close to the 1:1 line, and a few sources have higher interpolated fluxes than observed. 
This is expected, as we are only measuring the point-like component of the SCUBA-2 flux, while the large SPIRE beam ($\sim30\arcsec$) efficiently recovers flux that is extended to SCUBA-2.

\section{Sample overview and early science}\label{sec:obj}

\subsection{Spatial distribution of objects}\label{sec:galactocentric}

\begin{figure*}
    \centering
    \includegraphics[width=0.45\textwidth]{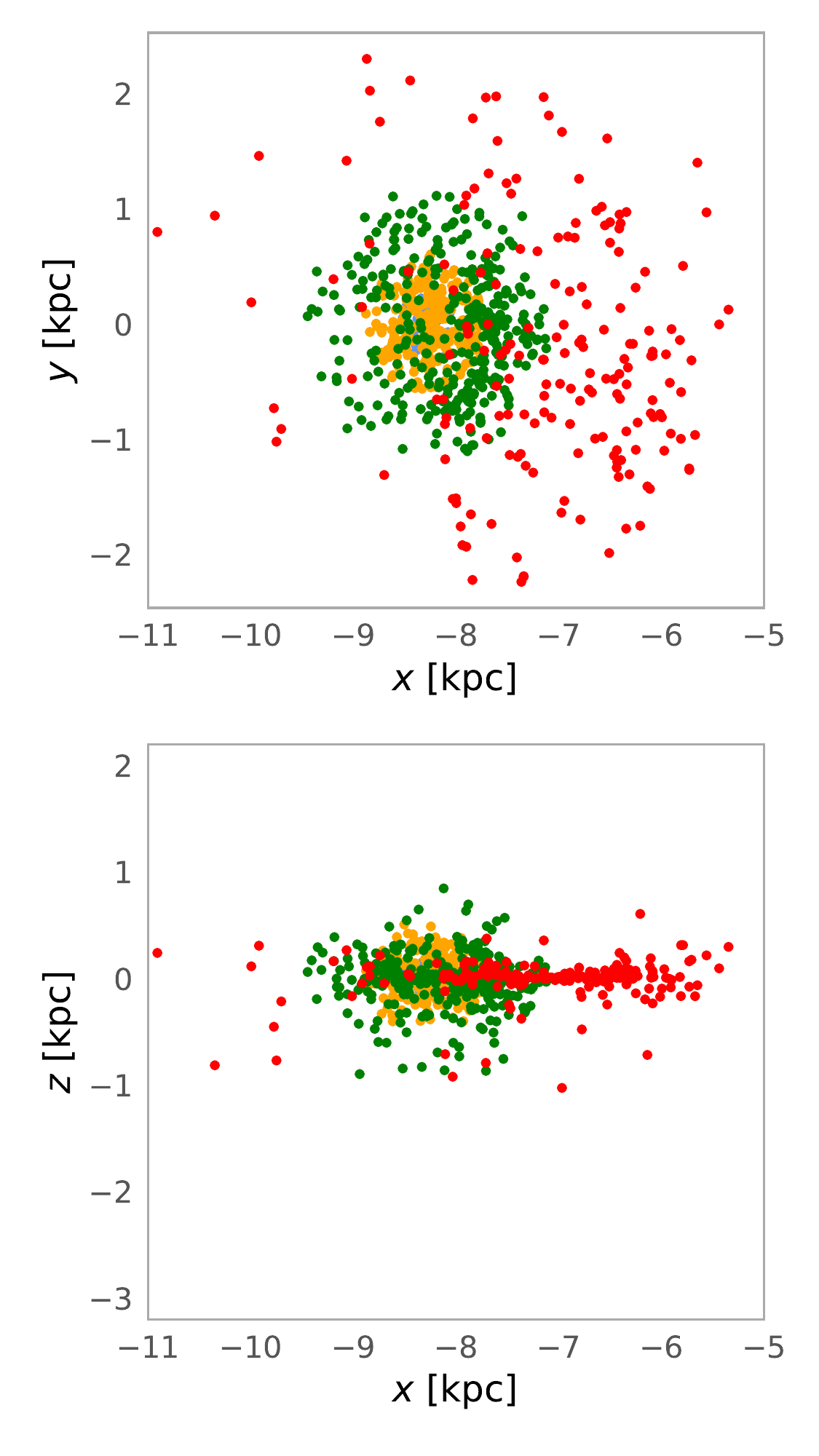}
    \includegraphics[width=0.45\textwidth]{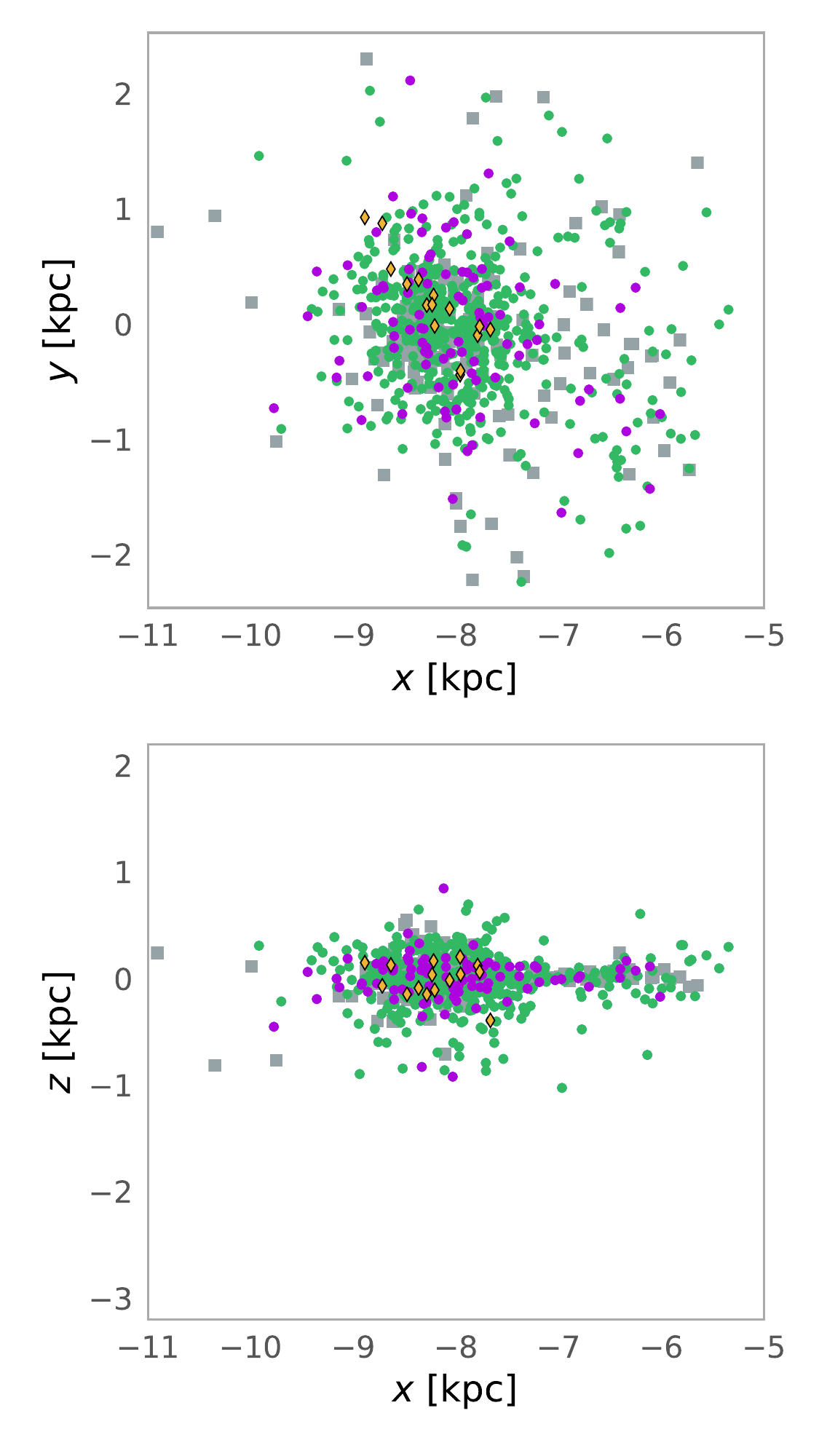}
    \caption{Left panels show the Galactocentric distribution of NESS sources, showing the NESS sample tiers (using the same colour scheme as Fig.~\ref{fig:sampledpr}). Right panels show the object chemistry: oxygen-rich stars are in green, carbon stars are in purple; S stars are in orange diamonds; stars without a spectroscopic chemical classification are in grey  squares.
    }
    \label{fig:samplesky_Galactocentric}
\end{figure*}
Figure~\ref{fig:samplesky_Galactocentric} shows the distribution of stars in Galacto-centric co-ordinates, coloured by their NESS tier and chemical type.

The ``extreme'' tier 4 (red in Figure~\ref{fig:samplesky_Galactocentric}) is dominated by populations in the plane of the Milky Way. These predominantly lie at Galactic radii interior to the Sun. This follows the distribution of hot main-sequence stars \citep[e.g.][]{Skowron19,Kounkel20}: regions interior to the Sun show significant star formation in the Sagittarius Arm, which merges into the Local Arm; exterior to the Sun, the Perseus Arm lacks much recent star formation. Recent star formation in these arms concentrates almost entirely within $\sim\pm$100 pc of the Galactic plane. Consequently, we expect tier 4 to comprise of massive AGB ($M_{\rm initial} \approx 4$\,--\,$8~ M_\odot$) and RSG ($M_{\rm initial} \gtrsim 8$ M$_\odot$) stars, from young ($\lesssim$150 Myr) populations.

Relatedly, tier 4 contains few carbon stars. This contrasts with metal-poor galaxies, where lower natal oxygen makes it easier for dredge-up to increase C/O to above unity \citep{KarakasLattanzio14}. In such galaxies, oxygen-rich stars remain significant dust producers, but carbon stars generally dominate the dust budget among extreme sources \citep[e.g.][]{Boyer12,Boyer17}. Tier 4 also includes any contaminating objects that made our conservatively inclusive cuts for AGB stars. These include stars exhibiting moderate interstellar extinction (which mimics the optical attenuation of circumstellar dust), possible young stellar objects, and other mis-classified objects. The distances of these sources are also the most uncertain, potentially causing their distribution to fan out at large radii from the Sun (e.g., near $x = -6$ kpc in Figure~\ref{fig:samplesky_Galactocentric}). A detailed analysis of their stellar parameters will need to be performed before we can properly ascertain the significance of these trends, and the relative scale heights of carbon stars, and massive and low-mass oxygen-rich stars.

The ``high'' mass-loss rate tier 3 (green) contains the largest fraction of the carbon-rich (purple) and S-type (orange) stars. These descend from stars of $\sim$2--3 M$_\odot$ ($\sim$0.4--1.7 Gyr in age; \citealt{Marigo17,Pastorelli19}). Compared to the youngest and most-extreme tier 4 sources, tier 3 sources are distributed at larger Galactic scale heights, consistent with the few $\times$ 100 pc expected for sources $\sim$1 Gyr in age \citep[e.g.][]{Kounkel20}. Tier 3 also contains the large majority of the Gould Belt: an oval-shaped region of star formation surrounding the Sun, and inclined relative to the Galactic Plane such that it reaches a height of 100 pc from the plane, which should contain a small number of massive, young, oxygen-rich evolved objects \citep[e.g.][]{McDonald17,Zari18}. However, we expect these objects to pass through the tier 3 stage of extremity relatively quickly, becoming tier 4 objects.

By virtue of the initial mass function, the ``intermediate'' and ``low'' mass-loss-rate tiers 2 and 1 (yellow and blue) once again tend to contain older, lower-mass and oxygen-rich stars, which do not reach the more extreme phases of tier 3. These are spherically distributed around the Sun, consistent with their large scale heights and lack of clustering due to recent star-formation.

\subsection{Integrated dust production}\label{sec:dpr}
\begin{figure}
    \centering
    \includegraphics[width=0.45\textwidth]{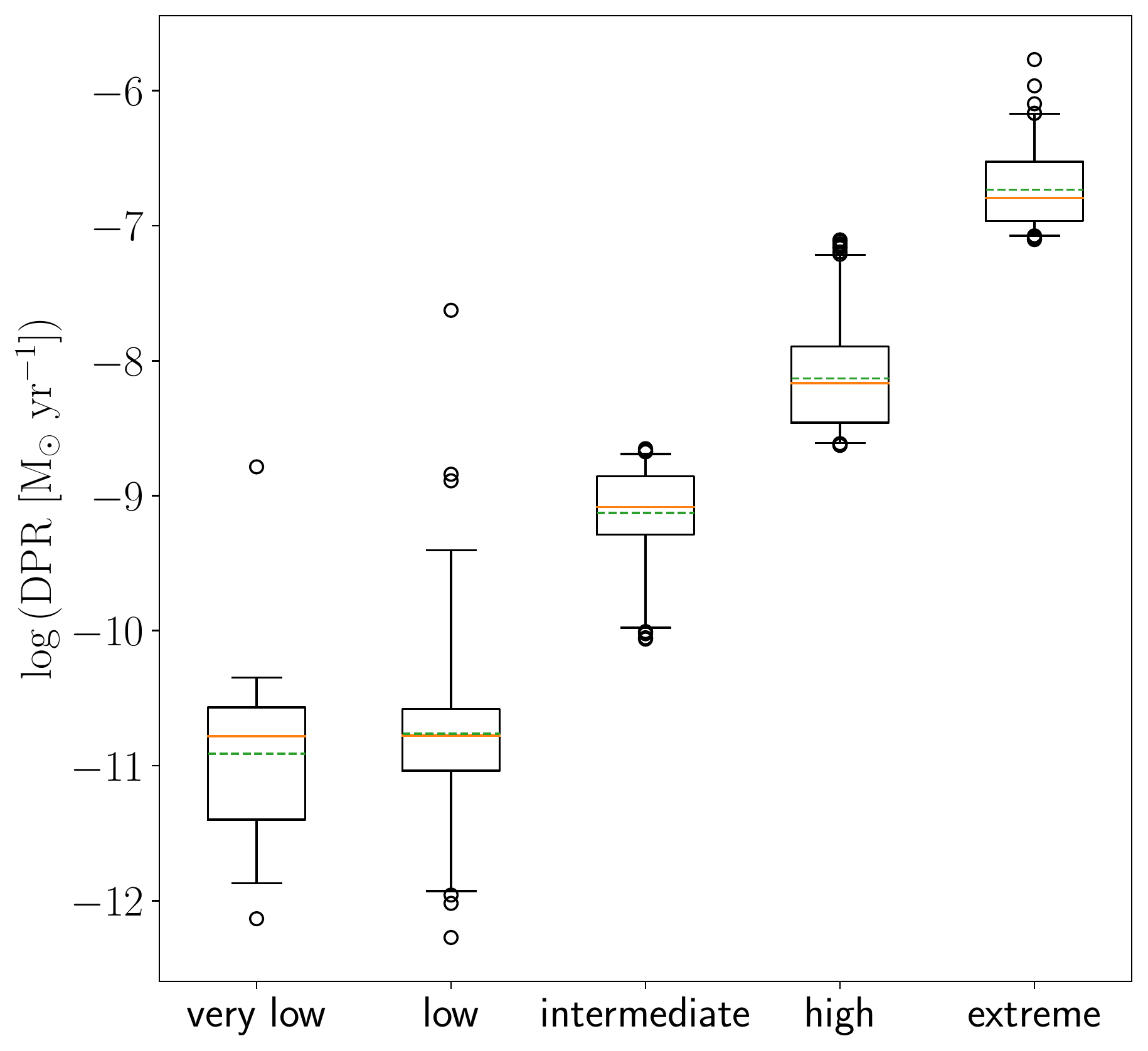}
    \caption{Box-and-whisker plots of preliminary dust-production rates in each of the five NESS tiers, as estimated from preliminary GRAMS model fitting. The boxes extend from the lower quartile to the upper quartile, enclosing the interquartile range (IQR; corresponding to the central 50 per cent of the data). The whiskers enclose the central 95 per cent of the data, and outliers are shown as open circles. The orange and green lines denote the median and mean values for each group.}
    \label{fig:dpr_dens}
\end{figure}

\begin{table}
    \caption{DPRs of the different tiers in the NESS sample.}\label{tab:dpr_dens}
    \centering
    \begin{tabular}{lcccc}
    \hline 
    Tier & No. & \multicolumn{3}{c}{DPR} \\
    & & Total &  Disc-averaged &  Volume-averaged\\
    & & M$_\odot$\,yr$^{-1}$& M$_\odot$\,yr$^{-1}$\,kpc$^{-2}$ & M$_\odot$\,yr$^{-1}$\,kpc$^{-3}$ \\
    \hline\hline
     Very low (0) &19&$1.9\times 10^{-9}$& $9.8\times 10^{-9}$& $2.9\times 10^{-8}$\\
     Low  (1) &105&$3.0\times 10^{-8}$& $1.0\times 10^{-7}$& $2.6\times 10^{-7}$\\
     Intermediate  (2) &222&$2.1\times 10^{-7}$& $1.9\times 10^{-7}$& $7.6\times 10^{-7}$\\
     High  (3) &324&$3.8\times 10^{-6}$& $8.3\times 10^{-7}$& $1.7\times 10^{-6}$\\
     Extreme  (4) &182&$4.3\times 10^{-5}$& $1.5\times 10^{-6}$& $1.9\times 10^{-5}$\\
    \hline
    Total  &852&$4.7\times 10^{-5}$& $2.6\times 10^{-6}$& $2.1\times 10^{-5}$\\
    \hline
    \end{tabular}
\end{table}

Figure \ref{fig:dpr_dens} shows the DPR distribution estimated from GRAMS fits to the SEDs of our sample stars. The tier design dictates that the mean DPR increases from the lowest tier to the highest. Tier 0 (``very low") targets were selected for their lack of dust production, hence their GRAMS DPRs should be treated as upper limits. The total DPR per unit volume in each tier is summarised in Table \ref{tab:dpr_dens}. For the first two tiers, the volumes are spheres with radii 250 and 300 pc, respectively. Extinction and confusion in the Galactic Plane ($|b| < 1.5$ deg) affects the more distant sources in tiers 2--4. For tiers 2 and 3 (``intermediate" and ``high"), we use spherical volumes of radii 400 and 800 pc, respectively, and add to this spherical volumes of radii 600 and 1200 pc, respectively, from which wedges of $|b| < 1.5$ deg have been removed. For tier 4 (``extreme"), we compute a cylindrical volume of radius 2000 pc with height 100 pc, corresponding to the Galactic scale height of their distribution (Figure \ref{fig:samplesky_Galactocentric}), and add to this the volume of a cylinder of radius 3000 pc after removing a wedge of $|b| < 1.5$ deg, as for the lower tiers. We also compute the surface densities for each tier for circles of radii 250, 300, 600, 1200, and 3000 pc, respectively. The total DPR for the Tier~1 sources is $3 \times 10^{-8}$ M$_\odot$ yr$^{-1}$, comparable to that of a single, dusty AGB star, while the total for  tier~2 is equivalent to the DPR of a single, extremely dusty AGB star.

Local dust production is dominated by the sources with the highest DPRs, comprising over half of the disc-integrated DPR of our entire sample. Table \ref{tab:dpr_dens} also indicates that these ``extreme'' sources comprise the vast majority of the volume-integrated DPR of our sample, but this is an artefact of the limited Galactic scale height used for computing this value, and only applies to the immediate 100 pc above and below the Galactic Plane. 
This implies that less-extreme stars should be more important at higher Galactic scale heights, with consequent changes to the fraction of carbon-rich dust produced, and the mineralogy of AGB ejecta overall.
However, the pronounced asymmetry in the distribution of Tier 4 sources with respect to the orbit of the Sun suggests a strong dependence of total dust production on Galactic radius and that evolved-star feedback is very inhomogeneous even on kpc scales. While the distribution of Tier 3 sources is also asymmetrical, the difference is much less significant than in Tier 4. This will have similar implications as their confinement to the Galactic Plane.

We remind the reader that our dust-production rates rely on assumptions that NESS sets out to test (e.g., the wind-velocity profile). However, they corroborate earlier estimates of Galactic dust production by AGB stars. \citet{Tielens10} estimates an integrated DPR of $8 \times 10^{-6}$ M$_\odot$ yr$^{-1}$ kpc$^{-2}$. Slightly lower rates were found by \citet{JuraKleinmann89} and \citet{Dwek98}. Our total rate (Table \ref{tab:dpr_dens}) is comparable to these rates, with the differences among the four publications consistent with different choices of dust emissivity and wind-acceleration profiles.

That mass loss is dominated by ``extreme'' stars is not surprising. \citet{LeBertre01} (also \citealt{LeBertre03}) found that half of mass loss can be attributed to stars with $\dot{M} > 10^{-6}$ M$_\odot$ yr$^{-1}$, despite the rarity (10 per cent) of such stars in their sample: assuming a gas-to-dust ratio of $\sim$200, their criterion corresponds to our tiers 3 and 4, so we tentatively identify an even larger fraction with these preliminary rates. Extreme stars also comprise most mass loss in nearby dwarf galaxies, however there the extreme stars are carbon-rich \citep[e.g.][]{Boyer12,Srinivasan16}. In the Milky Way, the higher metallicity and the consequent reduction of efficient dust production in the Milky Way by carbon stars with C/O $\gg$ 1, means we instead see most mass is lost by extreme \emph{oxygen}-rich objects (massive AGB stars, super-AGB and RSGs).

With all these statistics, however, we again caution that they are based on very preliminary DPRs. 
These stars also only reflect some of the Galaxy's AGB stars (albeit a large sample), and a fuller account will be given in Trejo et al.\ (in prep.).

\subsection{Spatially-resolved mass loss and irradiation of AGB envelopes}\label{sec:spatial}

At large radii from the star, CO becomes dissociated by interstellar UV light \citep{Mamon88,Groenewegen17,Saberi2019}. The extent of CO envelopes is an important input when modelling mass-loss rates of stars with unresolved envelopes. Measures of the local interstellar radiation field exist \citep{Habing68,Draine78,Mathis83}, but its variation and effects on the AGB envelopes are not well studied. Despite this, evidence is building that dissociation by strong UV radiation can significantly affect the recorded CO line strengths of stars \citep{McDonald15a,Groenewegen16,Li16,McDonald15b,McDonald19a}. 
Some previous studies \citep[e.g.][]{Olofsson90,Olofsson96,Castro-Carrizo2010} have resolved the CO envelopes of some stars in single-dish observations, typically in the (1--0) and sometimes (2--1) lines.

\begin{figure*}
    \centering 
    \includegraphics[width=0.475\textwidth]{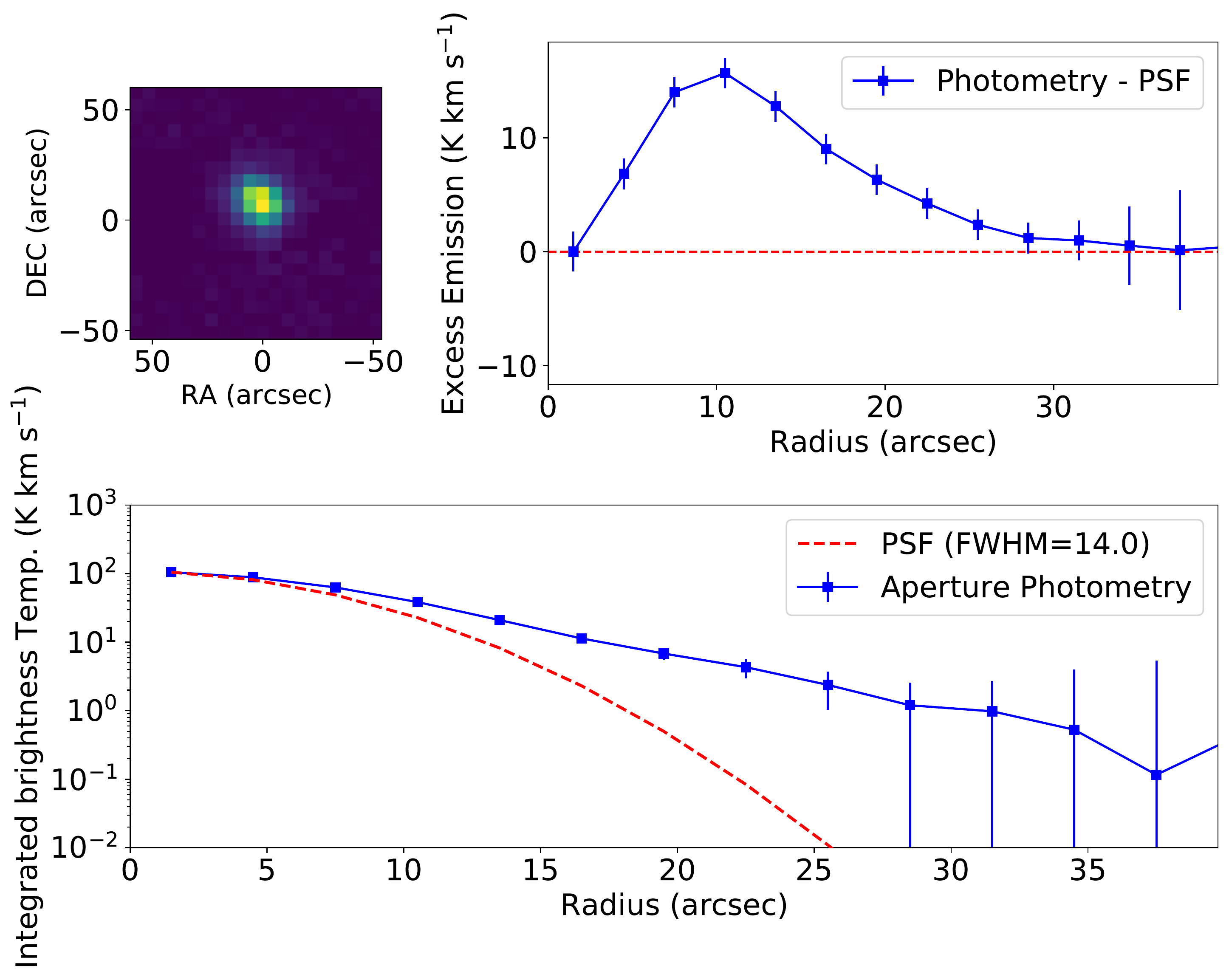}
    \includegraphics[width=0.475\textwidth]{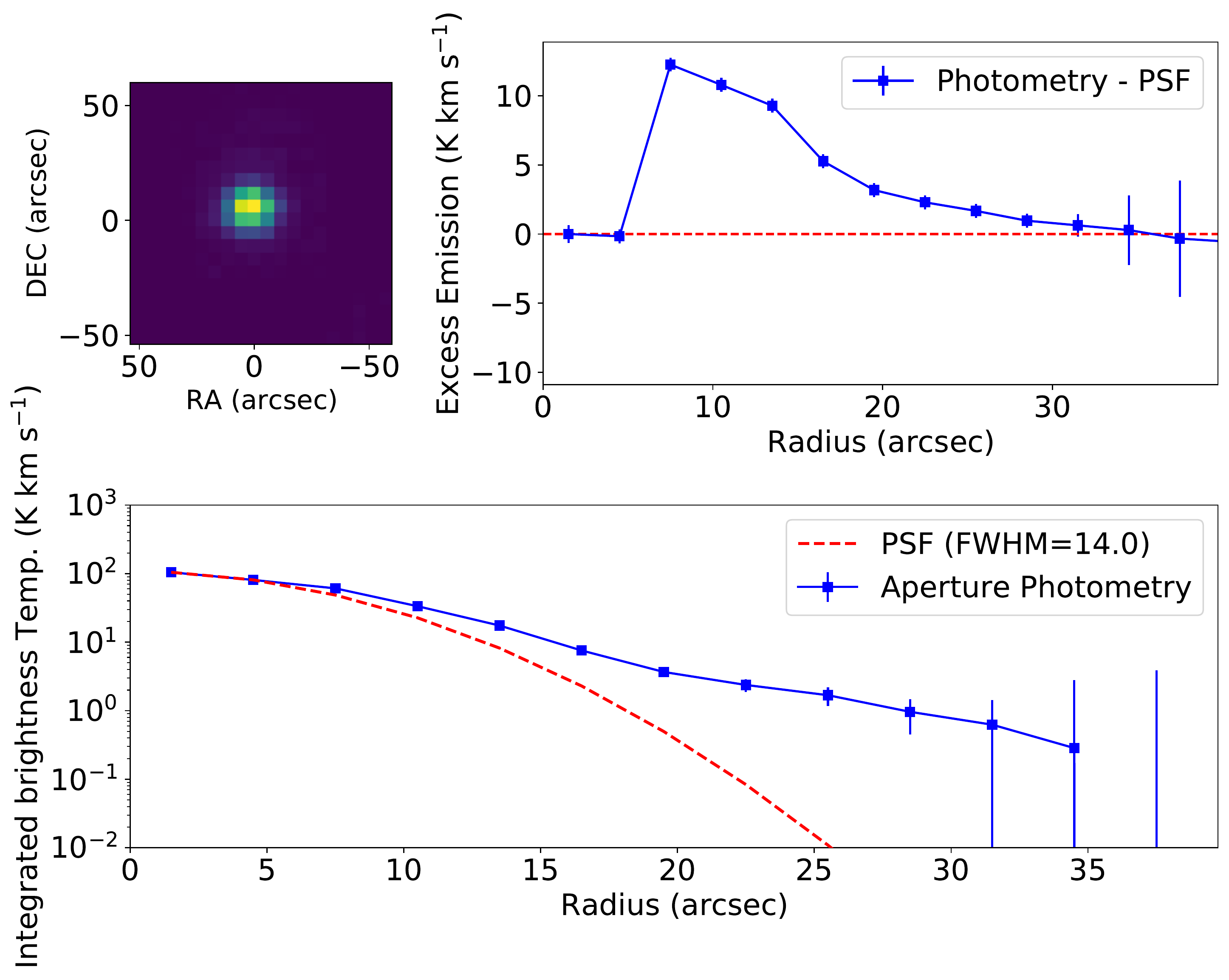}\\
    \includegraphics[width=0.475\textwidth]{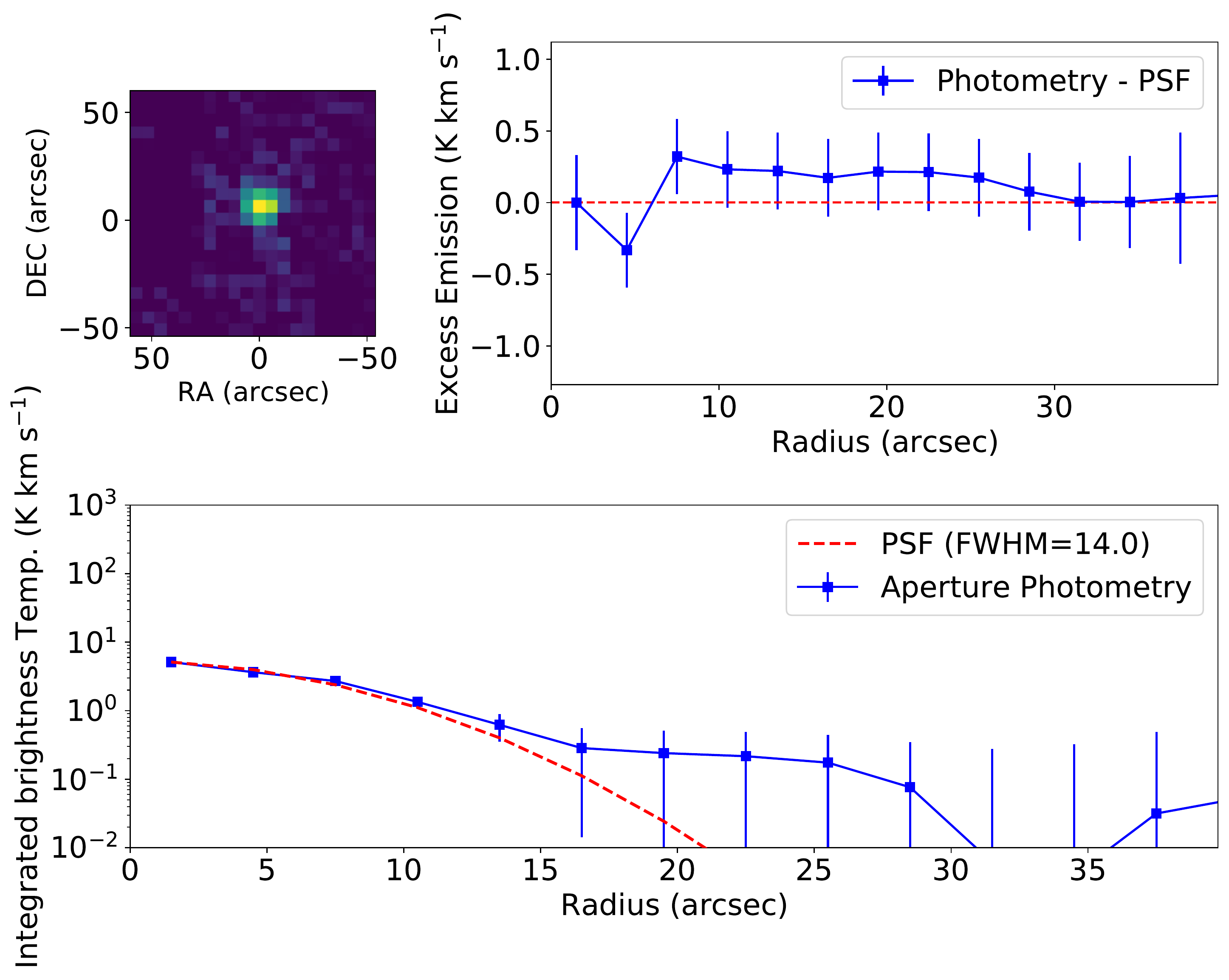}
    \includegraphics[width=0.475\textwidth]{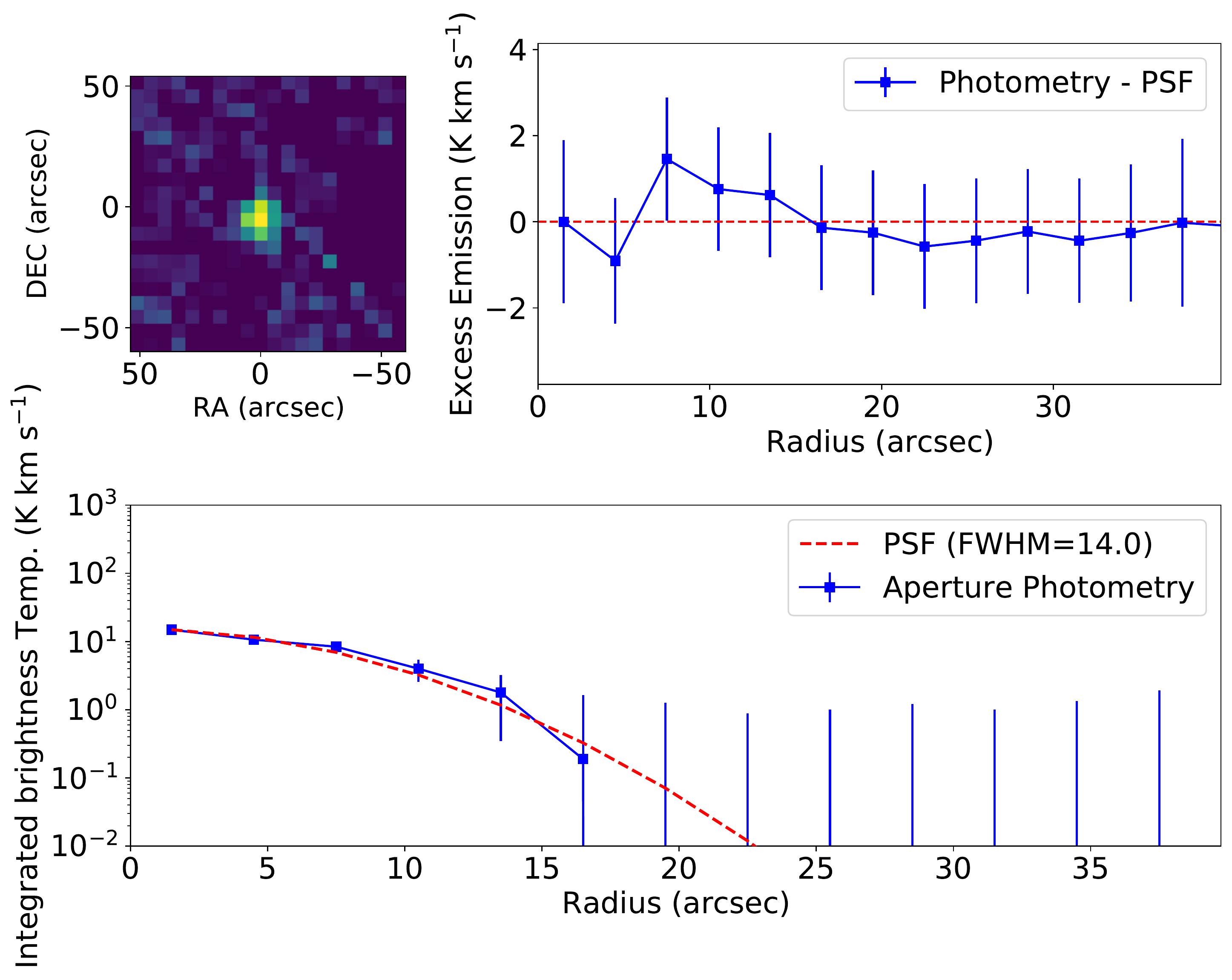}
    \caption{Examples of CO(3--2) radial profiles observed with the JCMT for four sources in the mapping sample. The top row shows sources that are clearly resolved {\it Left}: W Hya; {\it Right}: SW Vir. Extended CO emission is  clearly present in both sources at the level of $\sim10$ per cent of the peak flux. 
    The bottom row shows two unresolved sources {\it Left}: SV Lyn; {\it Right}: X Oph. The profiles match the 14\arcsec\ Gaussian assumed to be the FWHM beam width of the JCMT very well, validating our use of this to detect extended emission.
    }
    \label{fig:COradprof}
\end{figure*}

\citet{Dharmawardena2018} and \citet{Dharmawardena2019} have shown that coarse sub-mm mapping observations can also reveal the mass-loss history of the star.
The NESS mapping sample allows direct measurement of the extent of CO envelopes for a large sample of AGB stars, revealing the extended components of the CO(3--2) and (2--1) lines, providing an overview of the gas density and excitation in the outer envelope. 

Examples of the resulting data can be seen in Fig.~\ref{fig:COradprof}, showing both resolved and point-like sources.
The beam shapes of the current heterodyne instrumentation at the JCMT are poorly known , beyond the known $\eta_{\rm MB} \approx 0.6$. However, we expect a 15-metre telescope at 345\,GHz to have a Gaussian PSF with a FWHM of 14\arcsec. The dish surface is regularly adjusted using holography, and the amplitude of the error beam is roughly 2 per cent at 850\,\micron\ as seen by SCUBA-2 \citep{Dempsey13}. Given that the sub-mm seeing at Mauna Kea is typically $\lesssim2\arcsec$, the observations are clearly diffraction limited and atmospheric effects on the PSF are negligible.

The most-compact mapping sources all have radial profiles consistent with the expected PSF and its error beam, with no discernible variation between them and no evidence of flux leakage from sidelobes. Consequently, we can determine that the PSF does not change significantly between observations and any source observed to be more extended than these compact sources exhibits real extended emission.
Extended emission is present in the resolved sources in Fig.~\ref{fig:COradprof} at radii up to 30\arcsec, with the sizes and profiles of the emission region varying. 
This indicates that the emitting gas can reach a spatial extent of thousands of stellar radii, matching the prediction of \citet{Kemper03} that even relatively high-excitation lines such as CO(3--2) are excited sufficiently to be detectable at large distances from the central star. Using a preliminary set of CO maps from the JCMT, 
 we have measured the FWHM by fitting 1-D Gaussians to the radial profiles of the velocity-integrated line intensities from Fig.~\ref{fig:COradprof}. Of our sample of 25 mapping sources, 15 have a FWHM $> 18''$, making them clearly extended. Another 8 sources are marginally extended in CO(3--2) with $15'' < \mathrm{FWHM} < 18''$, while 2 sources, BK Vir and V744 Cen, remain completely unresolved for the JCMT beam, see
Tab.~\ref{tab:extCO}). This Table also shows the full extent of the CO(3--2) emission out to the 3-sigma detection limit. These results are 
comparable to the detection of extended sub-mm continuum emission in 14/15 (93 per cent) of evolved stars by \citet{Dharmawardena2018}, with the difference likely reflecting the inclusion of closer, lower $\dot{M}$ sources in the NESS mapping sample and lack of distant, high $\dot{M}$ objects.

\begin{table}
    \caption{ Extended CO(3-2) emission in our targets. The first two columns show the IRAS PSC identification and alternative names for our targets. The third column lists the FWHM determined by fitting a Gaussian to the radial profile of the velocity-integrated line intensity, while the fourth column shows the full extent of that quantity in the form of the diameter where the emission equals 3$\sigma$, averaged ringwise. Where no value is given this measurement could not be made, and instead we evaluate whether extended emission is marginal (?) or not detected (N) in the preliminary JCMT CO(3-2) maps. 
    Sources marked with $^\ast$ are also in the DEATHSTAR sample.} 
    \label{tab:extCO}
    \centering
    \begin{tabular}{llcc}
    \hline
    \emph{IRAS} PSC \# & {\sc SIMBAD} ID & FWHM ($''$) & Extent ($''$)\\\hline\hline
    IRAS 00205+5530     & T Cas&  15.2 & 27 \\
    IRAS 01556+4511     & V370 And  & 15.6 & 33 \\
    IRAS 03507+1115     & IK Tau &  19.2 & 57 \\
    IRAS 04020-1551     & V Eri &  17.4 & N \\
    IRAS 04566+5606     & TX Cam &  15.3 & 39 \\
    IRAS 07120-4433     & L$_2$ Pup$^\ast$  & 21.4 & 33 \\
    IRAS 08003+3629     & SV Lyn &  18.1 & N \\
    IRAS 09448+1139     & R Leo$^\ast$ &  21.6 & 51 \\
    IRAS 09452+1330     & IRC+10216  & 43.9 & 222  \\
    IRAS 10329-3918     & U Ant &  40.8 & 63 \\
    IRAS 11461-3542     & V919 Cen  & 15.7 & ? \\
    IRAS 12277+0441     & BK Vir$^\ast$ &  14.2 & N \\    
    IRAS 13001+0527     & RT Vir$^\ast$ &  19.0 & 33 \\
    IRAS 13114-0232     & SW Vir$^\ast$ &  18.4 & 51 \\
    IRAS 13368-4941     & V744 Cen  & 13.4 & ?\\
    IRAS 13462-2807     & W Hya$^\ast$  & 18.3 & 45 \\
    IRAS 15492+4837     & ST Her  & 18.8 & 27 \\
    IRAS 16235+1900     & U Her  & 20.2 & ? \\
    IRAS 18359+0847     & X Oph  & 15.0 & N \\
    IRAS 19039+0809     & R Aql  & 16.0 & 39 \\
    IRAS 19126-0708     & W Aql  & 18.0 & 63 \\
    IRAS 21088+6817     & T Cep  & 18.1 & 27 \\
    IRAS 21439-0226     & EP Aqr  & 15.6 & 45 \\
    IRAS 22196-4612     & $\pi_1$ Gru  & 23.3 & 51 \\
    IRAS 23320+4316     & LP And & 18.2 & 63 \\
         \hline
    \end{tabular}

\end{table}

Note that the FWHM and the outer extent of the CO(3--2) emission reported in Table~\ref{tab:extCO} differs from the values reported by \citet{Ramstedt20} for the DEATHSTAR sample. In this ALMA-observed sample, the FWHM obtained for 42 AGB stars, six of which overlap with the sources in Table~\ref{tab:extCO}), is reported to be between 1--10\arcsec. 
Such small structures however, are not detectable with the 14\arcsec\ beam of the JCMT at 345 GHz, and this level of detail will be smoothed out by the JCMT beam profile. 
Only in cases where the radial profile detected by ALMA actually roughly follows a Gaussian profile with a FWHM of 14\arcsec\ or more can we expect to see the same values between these two surveys. 
Additionally, due to the maximum-recoverable scale (MRS) of 18\arcsec, the DEATHSTAR ALMA observations are not able to measure extended emission beyond that size. 
Thus, it is evident that the ALMA observations obtained by the DEATHSTAR project, and the JCMT observations presented here are not directly comparable, but complement each other, and probe substructure and typical sizes on different size scales.  
The comparatively large JCMT primary beam combined with the lack of spatial filtering also leads to a higher surface-brightness sensitivity in our data, allowing us to detect faint emission at large distances from the central star. 
This is particularly clear in the case of SW Vir (see Fig.~\ref{fig:COradprof}, top right); emission is clearly visible on scales up to 30\arcsec, twice the photodissociation radius quoted by \citet{Ramstedt20} for this source, resulting in a diameter more than three times the size of their MRS. 
This highlights the particular advantage of NESS over interferometric studies: the larger map size and lack of spatial filtering enables us to recover emission on much larger scales.
Combining the high surface-brightness sensitivity with the volume-complete sample will enable minimally biased inference concerning how common such large envelopes are and under what conditions they can appear.

Resolved maps such as these can be used to determine the mass-loss history of these stars, plus trace large-scale asphericities in the outflows. Comparing dust continuum with CO line profiles also allows us to probe the dust:CO ratio, tracing CO-dark parts of the outflow and the excitation and dissociation of the CO molecules.

\subsection{Anomalous detection of cold dust}\label{sec:cold}

\subsubsection{Hidden cold dust reservoirs}\label{sec:colddust}

Deriving DPR and $\dot{M}$ from fitting optical--mid-infrared SEDs (Sect.~\ref{sec:tiers}) best traces material ejected from the star in the last few decades, as this hot and warm dust emits in the mid-infrared. However, it is relatively insensitive to variations in mass-loss rate or to reservoirs of material ejected in the past (Sect.~\ref{sec:comethods}).

Resolving the envelopes at longer wavelengths isolates emission from cold dust emitted millennia ago \citep[e.g.][]{vanderVeen1995,Dehaes2007,Ladjal2010,Dharmawardena2018}. 
\citet{Cox2012} notably found extended far-infrared (FIR) emission in 49 out of 78 evolved stars\footnote{Some of the 29 ``non-detections'' may have extended emission, as the authors were looking for specific shapes of the extended emission.}, mostly drawn from the Mass-loss of Evolved StarS (MESS) {\it Herschel} Guaranteed Time Key Programme \citep{Groenewegen2011}, confirming cold dust is commmonplace.

However, sensitivity to the coldest dust requires (sub-)mm observations.
\citet{Ladjal2010} found extended emission in four out of nine sources at 870\,$\mu$m using the APEX Large Bolometer Camera (LABOCA); and \citet{Dharmawardena2018} resolved sub-mm emission at one or more wavelengths in 14 out of 15 evolved stars with JCMT/SCUBA-2, with extended emission accounting for an average 40 per cent of the sub-mm flux. 
Furthermore, by deriving time-averaged mass-loss rates from the resolved emission, \citet{Dharmawardena2018} show that the dust mass in the resolved observations is typically a factor of a few larger than predicted by the GRAMS models, which assume a constant $\dot{M}$ out to a fixed radius.
This has potential to significantly increase the contribution of AGB stars to the interstellar dust reservoir, and/or their interaction with it, and allows us to probe longer-timescale variations in mass loss.

The deep SCUBA-2 mapping observations of NESS will more than double the existing sample of evolved stars with deep, wide-field imaging in the sub-mm, providing details on the range of contributions from extended emission.
Meanwhile, the shallower continuum observations of the wider sample will provide a statistical overview of cold dust across a range of classes of evolved stars.

The far-infrared spectral slope, $\alpha$ (where $F_\nu \propto \nu^\alpha$) provides the nature of the FIR emission source. In particular, it can identify variations in the dust-opacity slope, $\beta$ (where $\kappa_\nu \propto \nu^\beta$; $\alpha \approx -\left(2 + \beta\right)$ with some temperature dependence for very cold dust), which can depend on the composition and size of circumstellar dust grains \citep[e.g.][]{Hoogzaad2002,Dehaes2007}. Many authors have shown that the $\beta$ depends strongly on grain size \citep[e.g.][]{Kruegel1994,Rodman2006,Ricci2011,Testi2014} and that, assuming an MRN-like power-law size distribution, this dependence is strongest for maximum grain sizes $\lambda/2\pi < a_{\rm max} < 3\lambda$  \citep{Draine2006}. 
Our mapping observations are hence most sensitive to $a \sim 50~\micron - 3\,mm$, significantly larger than predicted by dust-formation models \citep[e.g.][]{Hofner2008} but consistent with the largest pre-solar grains \citep[e.g.][]{amari2014recent}.
Similarly, grains of different composition or structure can have very different spectral indices \citep[e.g.][]{Mennella1998,Jager1998}, with laboratory and theoretical values for typical dust analogues covering the range $\beta = 1 - 2.5$.
\citet{Dharmawardena2018} found variations in $\beta$ in the outer envelope of IRC+10216, suggesting evolution of the dust as it is processed by the interstellar radiation field and integrated into the ISM. 

\subsubsection{Mimics of cold dust}\label{sec:submmcontributors}

Other sources contributing to sub-mm emission include diffuse backgrounds, and the star's optical and radio photospheres. These can be disentangled by a combination of mapping observations and the spectral slope, $\alpha$.

Diffuse background emission is normally filtered, but inhomogeneities may add or subtract flux from the PSF core. Interstellar dust is expected to have $\alpha \approx -3.7$ \citep{PlanckCollaboration2014}, but bright patches will not normally be spatially coincident with the AGB star in mapping observations, and diffuse emission this strong and inhomogeneous is likely to affect only a few stars (notably those in the Galactic plane) and be visible on large scales.

The star's optical photosphere should have $\alpha = -2$, will be spatially compact, and is already included in the GRAMS model fit. Uncertainties in the GRAMS luminosity and temperature may lead to uncertainties in the extrapolated sub-mm flux, but will be small.

The star's radio photosphere (unresolved emission arising from free electrons interacting with H\,{\sc i} and H$_2$ in the stellar chromosphere) is opaque at millimetre wavelengths, with a wavelength-dependent radius \citep[e.g.][]{Matthews18}. These canonically have $\alpha \approx -1.86$ \citep{ReidMenten1997,Vlemmings19}: slightly shallower but insufficiently different to $\alpha = -2$ to distinguish from cold dust in the staring sample. The sub-mm photospheric radius is similar to the optical photosphere near the peak of the SED, but typically cooler ($\sim$2000 K), so the sub-mm flux does not greatly exceed that of the optical photosphere's Rayleigh--Jeans tail \citep[e.g.][]{Kervella16,Vlemmings19}.

Observations of IRC+10216 by \citet{Menten2006} also suggest the sub-mm contribution from both optical and radio photospheres is small: the radio flux can be extrapolated to the sub-mm following
$$F_{850} = \left(\frac{\nu}{350 \mathrm{GHz}}\right)^{-1.8} F_\nu \left(\frac{130 \mathrm{pc}}{D}\right)^2 \frac{6200 L_\odot}{L_{\rm IRC+10216}}$$
for frequency $\nu$, distance from the source $D$, and stellar luminosity $L_{\rm IRC+10216}$.
Given our detection limit of $\sim$ 11\,mJy, an X-band (8.4\,GHz) flux of $\sim 0.6$\,mJy (\citealt{Menten2006}, Table~2), and assuming that IRC+10216 is roughly twice as luminous as the median of the LMC population, only sources within $\sim$200\,pc would have detectable radio photospheric emission in our observations.

Stellar chromospheres can also contribute significant radio flux with a much shallower spectral dependence that can dominate stellar continuum emission at longer (radio) wavelengths \citep[e.g.][for Antares]{OGorman20}. The radio photosphere canonically extends to $\sim$2--3 stellar radii (R$_\ast$) at 22 or 43 GHz, and typically has a lower brightness temperature than the star's effective photospheric temperature \citep[e.g.][for Betelgeuse]{Lim98,OGorman15}. In the case of Betelgeuse, the chromosphere only contributes 10--20 per cent more flux than the expected Rayleigh--Jeans tail at 850\,\micron\ \citep{OGorman17}. Consequently, we expect little additional flux at these wavelengths from stellar chromospheres.

Other physical explanations for sub-mm excess include a low $\alpha$ due to variations in grain size and composition described above. However, mm-sized grains are not generally predicted to form in the outflow (the largest predicted and observed being a few $\mu$m in diameter; \citealt{Hofner2008, Norris12, Scicluna15b}), while few dust analogue materials have bulk spectral indices $\beta>-1$ \citep[e.g.][]{Mennella1998, Demyk17a, Demyk17b, Mutschke19}. 
\citet{Brunner2018A&A...614A..17B} investigated whether changes in grain structure, composition or size could explain similar excess emission in LABOCA observations of R Scl, but found all of these options unsatisfactory, ultimately suggesting that either additional dust components (such as polycyclic aromatic hydrocarbons) were required, or that the emissivities of the dust may exhibit a significant change in the sub-mm \citep[see also][in the Magellanic Clouds]{Gordon2014ApJ...797...85G}. While data reduction may also play a role, Fig.~\ref{fig:s2spire} suggests a good agreement between the NESS and interpolated SPIRE fluxes at 450\,$\mu$m: the only deviations being stars where we may \emph{underestimate} the flux by a factor of up to two, whereas anomalous cold dust will produce an excess of flux at both 450 and 850\,$\mu$m.

Consequently, we expect a marked excess in sub-mm flux in NESS should only be attributable to an anomalous cold dust reservoir, especially where the mapping sample resolves it from the point-source star.

\subsubsection{First NESS results}\label{sec:firstresults}

We here present some early results of NESS continuum observations, based on the data reduction and fluxes presented in Sect.~\ref{sec:red}.
At 850\,\micron\,our detection rate is 73 per cent, but varies significantly across our subsamples; the rate approaches 90 per cent for sources in Tier 2 (shown in green in Fig.~\ref{fig:scuba2fluxes}), while the detection rate in Tier 4 (in red) is just above 50 per cent.
The origin of this decrease is not yet clear, although increased confusion from diffuse interstellar dust at larger distances seems like an obvious contributor.  
Follow-up observations with a compact interferometer would reveal whether this is the case by filtering out the contamination. However, this requires care not to remove emission from historic dust mass loss.

The lower sensitivity at 450\,\micron\, leads to a reduced detection rate of $\sim$30 per cent.
A similar pattern exists throughout the subsamples, decreasing from $\sim 40$ per cent for Tier 2, to 24 per cent for Tier 4.

\begin{figure}
    \centering
    \includegraphics[width=0.475\textwidth, clip=true, trim=0.25cm 0.25cm 1.5cm 1.cm]{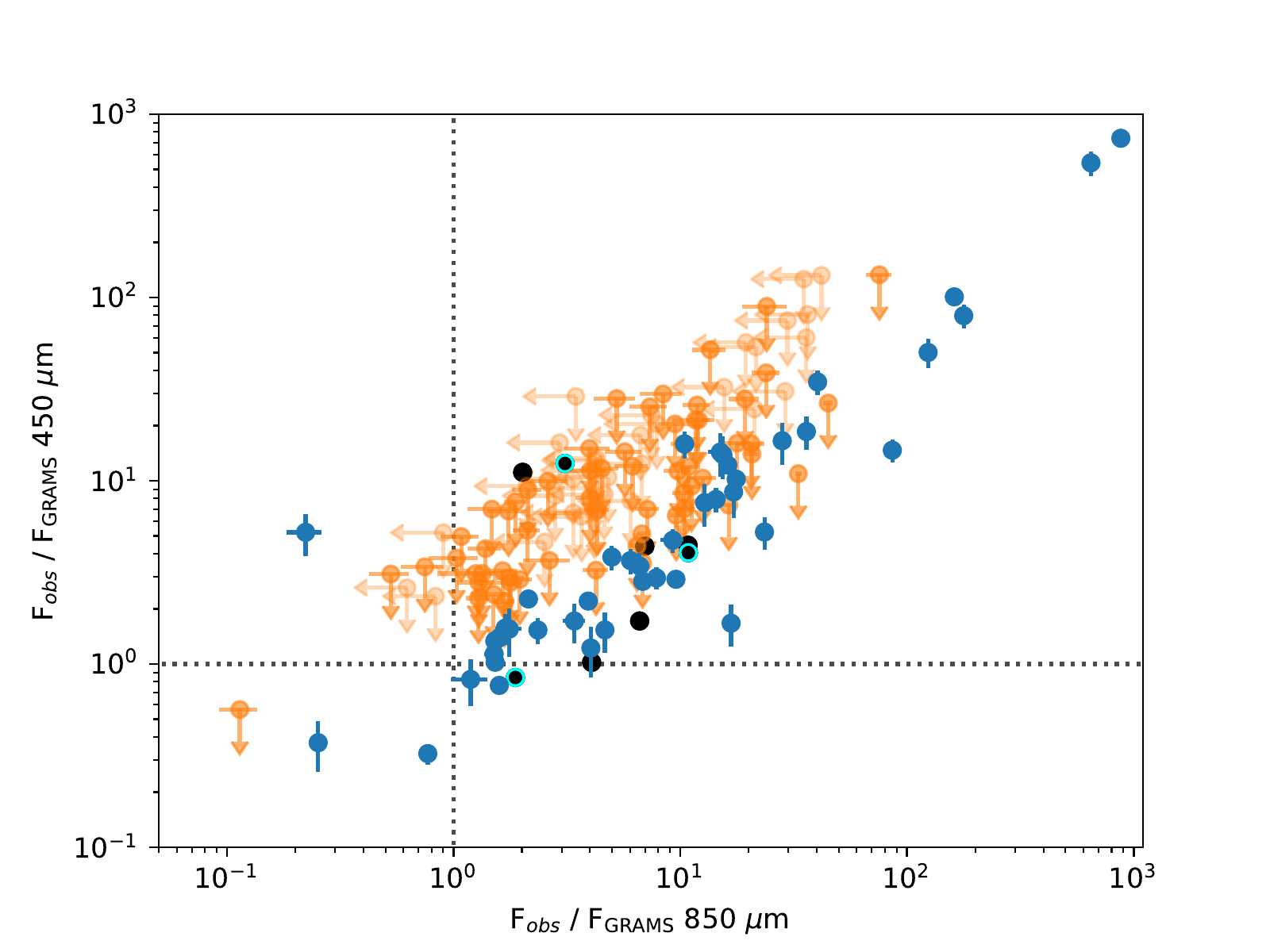}
    \caption{Ratio of SCUBA-2 fluxes to that predicted by the GRAMS fits for the subset of NESS sources shown in Fig.~\ref{fig:scuba2fluxes}, with blue points corresponding to sources detected at both wavelengths, and orange to sources detected at one or zero wavelengths. For these non-detections, the points indicate the ratio between the 3$\sigma$ upper limit and the GRAMS flux. Black points indicate sources where \citet{Dharmawardena2018} reported extended sub-mm continuum emission, and cyan points highlight those which also have extended CO emission in table~\ref{tab:extCO}.
    }
    \label{fig:scuba2excess}
\end{figure}

\begin{figure}
    \centering
    \includegraphics[width=0.475\textwidth, clip=true, trim=0.25cm 0cm 1.5cm 1.45cm]{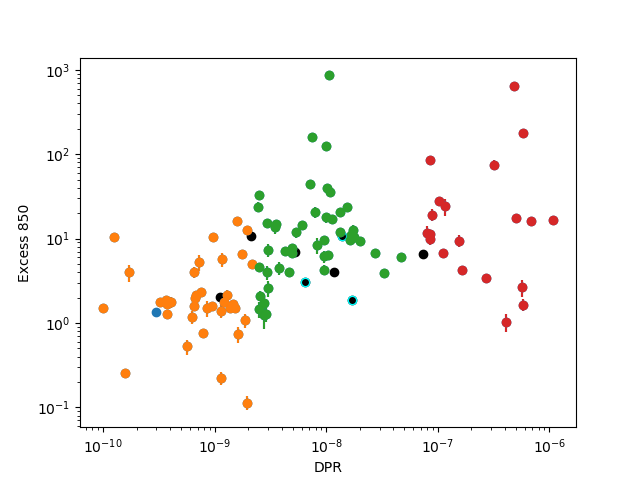}
    \caption{Ratio of observed 850\,\micron\ emission to that predicted by GRAMS fits as a function of dust-production rate, following the same colour code as Fig.~\ref{fig:sampledpr} (orange: Tier 2; green: Tier 3; red: Tier 4). As in Fig.~\ref{fig:scuba2excess}, black points indicate sources where \citet{Dharmawardena2018} reported extended sub-mm continuum emission, and cyan points highlight those which also have extended CO emission in table~\ref{tab:extCO}.}
    
    \label{fig:ExcessDPR}
\end{figure}

Fig.~\ref{fig:scuba2excess} compares our measured fluxes to the model predictions of GRAMS (Sect.~\ref{sec:tiers}). A 10 per cent calibration uncertainty has been added in quadrature to the statistical uncertainty.
The GRAMS models clearly and systematically underestimate the sub-mm continuum flux, typically by a factor of 3--10, the factor being larger at 850~\micron\ than 450~\micron.
This is not unexpected, given the limitations of the GRAMS grid (i.e., it is only intended to predict optical--mid-IR photometry and is tailored to the Magellanic Clouds), the uncertainties of the modelling (i.e., the models are fitted to relatively few bands in the near-/mid-infrared) and the uncertainties of the data reduction.
However, similar to \citet{Dharmawardena2018}, \citet{Dharmawardena2019} and \citet{Maercker2018}, we find systematically higher sub-mm fluxes than the models predict, arguing against data-reduction uncertainty as the origin of the discrepancy.
This might indicate that large reservoirs of cold dust are present, or that the dust properties are not well represented by those used in GRAMS.
This figure also includes sources where extended dust emission was detected by \citet{Dharmawardena2018}. These also all show systematic excess emission compared to the GRAMS models, hence physically linking a sub-mm excess flux (compared to GRAMS) with spatially extended cold dust emission.

Fig.~\ref{fig:ExcessDPR} plots the 850~$\mu$m excess as a function of the DPR derived from GRAMS,
roughly equivalent to plotting a mid-infrared colour against a mid-IR -- sub-mm colour (e.g. [12]$-$[25] vs.\ [25]$-$[850]).
A trend of increasing excess with DPR is visible, particularly across Tiers 2 and 3. To determine its statistical evidence, we use Bayesian model selection to calculate the posterior odds \citep{KassRaffertyBayesFactors}, using {\sc dynesty} to compute the evidence with Nested sampling \citep{SkillingNS2004AIPC..735..395S, dynestySpeagle2020MNRAS.493.3132S}, following \citet{Dharmawardena2020}. 
This approach shows that a power-law relationship is preferred over no relationship by a significant margin ($\sim 10^{10\,000}$), while a power law with a constant flux above a break point is indistinguishable from the unbroken power law.
While this framework is susceptible to some uncertainty, particularly from dependence on the choice of prior, it is difficult to envisage this overcoming a multi-googol preference.

The easiest interpretation of this trend relates to the amount of cold dust in the envelopes:
stars with higher present-day DPRs will tend to have been producing dust for longer, and are therefore likely to have filled a greater fraction of the SCUBA-2 beam with dust. Due to specific circumstances of source distances the cold dust component does not have to be extended (spatially resolved), but by the dictates of physics the extended component has to be cold (assuming thermal equilibrium).
As the GRAMS models use a fixed-density profile, dust composition and outer radius, they do not account for physical changes that may impact the sub-mm emission; detailed models which treat the sub-mm fluxes and radial profiles appropriately may better capture the behaviour. 
Stochastic changes in DPR or differences in dust composition may be responsible for the scatter.

\begin{figure}
    \centering
    \includegraphics[width=0.475\textwidth, clip=true, trim=0.5cm 0.25cm 1.5cm 1.45cm]{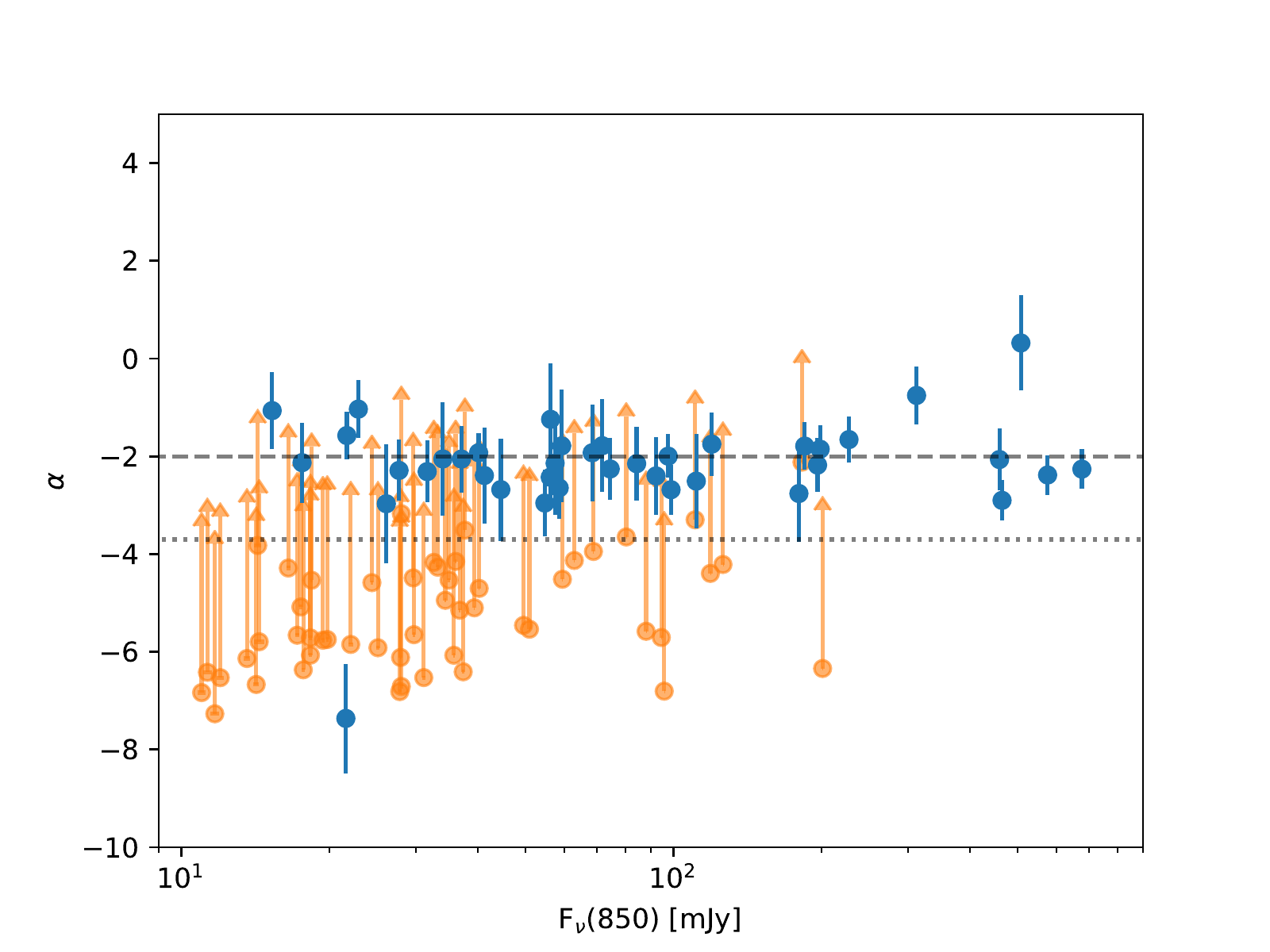}
    \caption{Sub-mm spectral indices for sources shown in Fig.~\ref{fig:scuba2fluxes}. Blue points with error bars indicate sources whose spectral index is well-constrained, and the orange points are lower limits, indicating sources that are detected at 850~$\mu$m but not 450~$\mu$m. The black dashed line corresponds to the result expected for blackbody emission, and the dotted line to the canonical value of --3.7 expected for interstellar dust \citep{PlanckCollaboration2014}. 
    }
    \label{fig:scuba2alpha}
\end{figure}

We derive spectral indices from the above SCUBA-2 fluxes (Fig.~\ref{fig:scuba2alpha} and Tab.~\ref{tab:flux}), ensuring that the results are robust and upper limits are handled consistently, by adopting the method of \citet{Scicluna2020aMNRAS.494.2925S}, to which we refer the reader for the details of the fitting and parameters for the  Markov Chain Monte Carlo (MCMC).
As expected, sources with a well-constrained spectral index (blue points) separate from lower limits (orange); the latter corresponding to sources detected at 850~$\mu$m but not 450~$\mu$m, which can have any positive spectral index.
In general, the upper limits at 450~$\mu$m do not place strong constraints on $\alpha$, though the limits remain consistent with the rest of the sample.
Sources with well-constrained $\alpha$ cluster around $\alpha \approx -2$, which manifests as the tight correlation in Figure~\ref{fig:scuba2excess}, and consistent with a cold-dust reservoir (Sec.~\ref{sec:submmcontributors}).

The two outliers (one at positive $\alpha$ and one at negative $\alpha$) are also the two outliers in Fig.~\ref{fig:scuba2fluxes}.
The source at $\alpha=0.3\pm1.0$ is IC\,418, a planetary nebula with significant free-free emission. The spectrum continues to rise into the radio, 1.72 Jy at 6\,cm \citep{Griffith1994}.
The source at $\alpha=-7.4\pm1.1$ is AK~Hya, a nearby AGB star with a relatively low mass-loss rate and high space motion \citep[cf.][]{McDonald18}.
To eliminate the discrepancy, the true 450~$\mu$m flux would have to be lowered by factor of 20, which seems improbable even for a 4-$\sigma$ detection, as the expected frequency of such outliers is $\sim 0.00007$.
If real, the origin of such an anomalous spectral index is an enigma; a small number of nearby debris discs have similarly steep spectra, attributed to anomalous dust compositions or size distributions \citep[e.g.][]{Ertel2012, Marshall2016}.

\section{Summary}\label{sec:sum}
We have presented the ongoing Nearby Evolved Stars Survey, representing a volume-complete sample of 852 Galactic evolved stars within 3~kpc, suitable for statistically robust inference of the properties of the population.
We discussed the observing strategy and the survey's key scientific objectives, and introduced the public NESS catalogue of 852 stars, which will be populated with new data as it becomes available. We introduce an improved distance-estimation method for AGB stars, based on the LMC luminosity distribution, and accurate to $\pm\sim$25 per cent, and anchor this to \emph{Gaia} EDR3 parallaxes.

The sample covers both low- and high-mass AGB stars. 
Objects with the highest DPRs concentrate in young populations near the Galactic plane; hence, are likely massive AGB stars and RSGs. Older populations, with large Galactic scale heights, have systematically lower DPRs.
Similar to previous studies, a few sources with the highest DPRs dominate the overall dust production by evolved stars; however, in contrast to the well-studied local dwarf galaxies, the dominant dust-producers in our sample are oxygen-rich. We fit models to the SEDs of our sampled stars to estimate their DPRs, resulting in an integrated value of $4.7\times 10^{-5}$ M$_\odot$ yr$^{-1}$ for the entire sample (Sect.~\ref{sec:dpr}). We anticipate these can be improved by incorporating more archival photometry.

Our resolved CO(3--2) observations, show significant extended emission on scales of $\sim$30\arcsec\ across many of the sources we selected for mapping (Sect.~\ref{sec:spatial}).
These data can be used to explore the gas mass-loss history, complementary to existing results on the dust mass-loss history \citep[e.g.][]{Dharmawardena2018},
revealing whether the dust-to-gas ratio varies throughout the circumstellar envelope.

An initial analysis of sub-mm continuum observations from the JCMT shows that the sub-mm emission of evolved stars is generally consistent with blackbody emission in the Rayleigh--Jeans regime (Sect.~\ref{sec:cold}), though two sources stand out for having very shallow (IC\,418) or steep (AK\,Hya) spectral indices.
However, the remaining fluxes are up to a factor of 10 higher than the predictions of radiative-transfer models fitted to shorter wavelengths.
Comparison with previously publications of spatially resolved dust emission suggests that unexpected large reservoirs of cold dust are present, though we cannot rule out that the properties of the dust are also not well represented by our models.

While a large part of the NESS sample will most likely be too bright for observations with future, sensitive facilities such as the extremely large class of telescopes, \emph{James Webb Space Telescope} or the Vera C.\ Rubin Observatory (formerly LSST), the detailed studies of these nearby objects will be necessary to inform studies of more distant sources with those future facilities.
The well-constrained dust-to-gas ratios will prove useful to interpreting observations of resolved extra-galactic populations.

\section*{Acknowledgements}
Wayne Holland sadly passed away while this paper was being prepared, but not before making a huge contribution to the paper and the NESS project as a whole. We wish to dedicate this paper to his memory. Wayne was a great scientist, mentor and friend who will be sorely missed.

We would like to thank the referees for their quick and constructive feedback which has helped to improve this manuscript.

The James Clerk Maxwell Telescope is operated by the East Asian Observatory on behalf of The National Astronomical Observatory of Japan; Academia Sinica Institute of Astronomy and Astrophysics; the Korea Astronomy and Space Science Institute; Center for Astronomical Mega-Science (as well as the National Key R\&D Program of China with No. 2017YFA0402700). Additional funding support is provided by the Science and Technology Facilities Council of the United Kingdom and participating universities in the United Kingdom and Canada. Program ID: M17BL002.

This research has been financially supported by the Ministry of Science and Technology of Taiwan under grant numbers MOST104-2628-M-001-004-MY3 and MOST107-2119-M-001-031-MY3, and by Academia Sinica under grant number AS-IA-106-M03. IM acknowledges support from the UK Science and
Technology Facilities Council under grants ST/L00768/1 and ST/P000649/1.
IM and AAZ acknowledge support from the UK Science and Technology Facility Council under grants ST/L000768/1 and ST/P000649/1.
SS acknowledges support from UNAM-PAPIIT Programme IA104820. SHJW acknowledges support from the Research Foundation Flanders (FWO) through grant 1285221N, and the ERC consolidator grant 646758 AEROSOL.
OCJ has received funding from the EU's Horizon 2020 programme under the Marie Sklodowska-Curie grant agreement No 665593 awarded to the STFC. 
JThvL, HI and PS were supported by Daiwa Anglo-Japan Foundation and the Great Britain Sasakawa Foundation. 
HC and HLG acknowledge support from the European Research Council (ERC) in the form of Consolidator Grant CosmicDust (ERC-2014-CoG-647939).
MM acknowledges support from an STFC Ernest Rutherford fellowship (ST/L003597/1).
JC and EP are support by a Discovery Grant from the Natural Sciences and Engineering Research Council (NSERC).
JHH thanks the National Natural Science Foundation of China under grant Nos. 11873086 and U1631237 and support by the Yunnan Province of China (No.2017HC018).
JPM acknowledges support from the Ministry of Science and Technology of Taiwan under grant number MOST109-2112-M-001-036-MY3.
GR acknowledge that the material is based upon work supported by NASA under award number 80GSFC17M0002.
H.K. acknowledges support by the National Research Foundation of Korea (NRF) grant funded by the Korea government (MIST) (No. 2021R1A2C1008928).
This work is sponsored (in part) by the Chinese Academy of Sciences (CAS), through a grant to the CAS South America Center for Astronomy (CASSACA) in Santiago, Chile.
This research has made use of the VizieR catalogue access tool, CDS,
Strasbourg, France (DOI : 10.26093/cds/vizier). The original description 
of the VizieR service was published in A\&AS 2000 143, 23.

{
\section*{Data Availability}

The data underlying this article are available in a number of ways. 
All raw JCMT observations are stored in the CADC archive at \url{https://www.eaobservatory.org/jcmt/science/archive/} and can be accessed under program IDS M15BP047 and M17BL002.
Catalogues containing all data compiled for or produced in this paper are available at \url{https://evolvedstars.space}.
 }



\bibliographystyle{mnras}
\bibliography{biblio.bib}





\vspace{0.5cm}
\noindent

\section*{Author affiliations}

$^{1}$Institute of Astronomy and Astrophysics, Academia Sinica, 11F of AS/NTU Astronomy-Mathematics Building, No.1, Sec. 4, Roosevelt Rd, Taipei 10617, Taiwan\\
$^{2}$European Southern Observatory, Alonso de Cordova 3107, Santiago RM, Chile\\
$^{3}$European Southern Observatory, Karl-Schwarzschild-Str. 2, 85748 Garching b. M\"unchen, Germany \\
$^{4}$Jodrell Bank Centre for Astrophysics, Department of Physics and Astronomy, School of Natural Sciences, University of Manchester, M13 9PL, Manchester, UK \\
$^{5}$Department of Physical Sciences, The Open University, Walton Hall, Milton Keynes, MK7 6AA, UK\\
$^{6}$Instituto de Radioastronom\'ia y Astrof\'isica, UNAM. Apdo. Postal 72-3 (Xangari), Morelia, Michoac\'an 58089, Michoac\'{a}n, Mexico\\
$^{7}$Institute of Astronomy, KU Leuven, Celestijnenlaan 200D bus 2401, 3001 Leuven, Belgium\\
$^{8}$East Asian Observatory (JCMT), 660 N. A`ohoku Place, Hilo, Hawai`i, USA, 96720\\
$^{9}$Department of Physics and Astronomy, University of Western Ontario, London, ON, N6A 3K7, Canada\\
$^{10}$Institute for Earth and Space Exploration, University of Western Ontario, London, ON, N6A 3K7, Canada\\
$^{11}$SETI Institute, 189 Bernardo Avenue, Suite 100, Mountain View, CA 94043, USA\\
$^{12}$School of Physics \& Astronomy, Cardiff University, The Parade, Cardiff CF24 3AA, UK\\
$^{13}$Yunnan Observatories, Chinese Academy of Sciences, 396 Yangfangwang, Guandu District, Kunming, 650216, China \\
$^{14}$Chinese Academy of Sciences South America Center for Astronomy, National Astronomical Observatories, CAS, Beijing 100101, China \\ 
$^{15}$Departamento de Astronom\'ia, Universidad de Chile, Casilla 36-D, Santiago, Chile\\
$^{16}$Department of Astrophysics, Vietnam National Space Center (VNSC), Vietnam Academy of Science and Technology (VAST), 18 Hoang Quoc Viet, Cau Giay, Ha Noi, Viet Nam\\
$^{17}$Graduate University of Science and Technology (GUST), Vietnam Academy of Science and Technology (VAST), 18 Hoang Quoc Viet, Cau Giay, Ha Noi, Viet Nam \\
$^{18}$Korea Astronomy and Space Science Institute (KASI) 776, Daedeokdae-ro, Yuseong-gu, Daejeon 34055, Republic of Korea \\
$^{19}$UK Astronomy Technology Centre, Royal Observatory, Blackford Hill, Edinburgh, EH9 3HJ, UK \\
$^{20}$Department of Physics and Astronomy, Kagoshima University, 1-21-35 Korimoto, Kagoshima, Japan \\
$^{21}$Amanogawa Galaxy Astronomy Research Center (AGARC), Graduate School of Science and Engineering, Kagoshima University, 1-21-35 Korimoto, Kagoshima 890-0065, Japan \\ 
$^{22}$Space Telescope Science Institute, 3700 San Martin Drive, Baltimore, MD 21218, USA \\
$^{23}$Max-Planck-Institute for Astronomy, K\"onigstuhl 17, 69117 Heidelberg, Germany. \\
$^{24}$Center for General Education, Institute for Comprehensive Education,  Kagoshima University, 1-21-30 Korimoto, Kagoshima 890-0065, Japan \\ 
$^{25}$Lennard-Jones Laboratories, Keele University, ST5 5BG, UK \\
$^{26}$Max-Planck-Institut f{\" u}r Radioastronomie, Auf dem H{\" u}gel 69, 53121 Bonn, Germany\\
$^{27}$Department of Physics and Astronomy, University College London, Gower St, London WC1E 6BT, UK \\
$^{28}$National Astronomical Observatory of China, Datun Road 20, Chaoyang, Beijing, 100012, China\\
$^{29}$Department of Earth Sciences, National Taiwan Normal University, Taipei 11677, Taiwan \\
$^{30}$CAS Key Laboratory of FAST, National Astronomical Observatories, Chinese Academy of Science, Beijing 100101, China\\
$^{31}$Department of Physics \& Astronomy, Texas Tech University, Box 41051, Lubbock TX 79409-1051, USA\\
$^{32}$Center for Astrophysics | Harvard \& Smithsonian, Smithsonian Astrophysical Observatory, USA\\
$^{33}$Department of Physics \& Astronomy, McMaster University, Hamilton, ON L8S 4M1, Canada\\
$^{34}$Department of Physics and Atmospheric Science, Dalhousie University, Halifax, NS B3H 4R2, Canada\\
$^{35}$Shanghai Astronomical Observatory, Chinese Academy of Sciences, 80 Nandan Road, Shanghai 200030, China\\
$^{36}$Xinjiang Astronomical Observatory, Chinese Academy of Sciences, 150 Science 1-Street, 830011 Urumqi, China\\
$^{37}$INAF, Osservatorio Astronomico di Roma, Via Frascati 33, 00077 Monte Porzio Catone (RM), Italy \\
$^{38}$Institut f\"ur Theoretische Astrophysik, Zentrum f\"ur Astronomie der Universit\"at Heidelberg, Albert-\"Uberle-Strasse 2, D-69120 Heidelberg, Germany\\
$^{39}$Okayama Branch Office, Subaru Telescope, NAOJ, NINS 3037-5 Honjo, Kamogata, Asakuchi, Okayama, 719-0232, Japan \\
$^{40}$Department of Astronomy and Institute of Theoretical Physics and Astrophysics, Xiamen University, China\\
$^{41}$Centre for Astronomy, School of Physics, National University of Ireland Galway, Galway H91 CF50, Ireland \\
$^{42}$Institute for Scientific Research, Boston College, 140 Commonwealth Avenue, Chestnut Hill, MA 02467, USA \\
$^{43}$Universit\'e C\^ote d'Azur, Observatoire de la C\^ote d'Azur, CNRS, Lagrange, France\\
$^{44}$National Astronomical Observatories, Chinese Academy of Sciences, Beijing 100101, China \\ 
$^{45}$Key Laboratory of FAST, National Astronomical Observatories, Chinese Academy of Sciences, Beijing 100101, China\\
$^{46}$School of Astronomy and Space Science, University of Chinese Academy of Sciences, Beijing 101408, China \\
$^{47}$Sterrenkundig Observatorium, Universiteit Gent, Krijgslaan 281 S9, B-9000 Gent, Belgium\\
$^{48}$Department of Physics, The University of Hong Kong, Pok Fu Lam Road, Hong Kong SAR, China \\
$^{49}$Laboratory for Space Research, The University of Hong Kong, Pok Fu Lam Road, Hong Kong SAR, China\\
$^{50}$Institute of Cosmology and Gravitation, University of Portsmouth, Burnaby Road, PO1 3FX, Portsmouth, UK\\
$^{51}$Centre for Astrophysics, University of Southern Queensland, West Street, Toowoomba, QLD 4350, Australia\\
$^{52}$Department of Astronomical Sciences, The Graduate University for Advanced Studies (SOKENDAI), 2-21-1 Osawa, Mitaka, Tokyo 181-8588, Japan\\ 
$^{53}$Okayama Observatory, Kyoto University, Kamogata, Asakuchi, Okayama 719-0232, Japan \\
$^{54}$School of Astronomy and Space Science, Nanjing University, Nanjing, 210093, China \\
$^{55}$NASA Goddard Space Flight Center, 8800 Greenbelt Road, Greenbelt, MD 20771, USA\\
$^{56}$Department of Physics, The Catholic University of America, Washington, DC 20064, USA\\
$^{57}$School of Sciences, Hainan University, Hainan 570228, China \\
$^{58}$Key Laboratory of Modern Astronomy and Astrophysics, Nanjing University, Ministry of Education, Nanjing 210093, China \\
$^{59}$Department of Physics and Astronomy, University of North Carolina, Chapel Hill, NC 27599-3255, USA \\
$^{60}$Kuffner Observatory, Johann-Staudstrasse 10, 1160, Vienna, Austria \\
$^{61}$Institute of Space and Astronautical Science, JAXA, 3-1-1 Yoshinodai, Chuo-ku, Sagamihara, Kanagawa 252-5210, Japan\\ 
$^{62}$Department of Space and Astronautical Science, SOKENDAI, 3-1-1 Yoshinodai, Chuo-ku, Sagamihara, Kanagawa 252-5210, Japan \\
$^{63}$School of Physics and Astronomy, Sun Yat-sen University, 2 Daxue Road, Tangjia, Zhuhai, China \\

\bsp	
\label{lastpage}
\end{document}